# Scaling of Low-Temperature Heat Capacity in Cryocrystals

M. Barabashko[1*], A. Jeżowski[2], A. Krivchikov[1,2+]
[1] *B.Verkin Institute for Low Temperature Physics and Engineering of the National Academy of Sciences of Ukraine, 47 Nauka Ave., 61103 Kharkov, Ukraine*

[2] *Institute of Low Temperature and Structure Research PAS, Okólna 2, 50-422 Wrocław, Poland*
[*]*ORCID:* https://orcid.org/0000-0003-3168-7119
[+]*ORCID:* https://orcid.org/0000-0001-5375-439X
*barabaschko@ilt.kharkov.ua*

## Abstract

The low-temperature isochoric heat capacity of cryocrystals was scaled the universal scaling function. This universality links the magnitude of the anomaly and the characteristic temperature of hump Tmax in heat capacity that related with the first van Hove singularity in the the phonon spectrum. For atomic, molecular and quantum cryocrystals, Tmax systematically shifts with molar volume, reflecting Brillouin zone scaling and revealing a common vibrational origin. These findings for scaling function is bridging thermodynamics with the vibrational density of states, and highlighting fundamental universality in lattice dynamics.

## 1. Introduction

The low-temperature heat capacity of solids reflects the quantum nature of atomic vibrations and the lattice dynamics. Cryocrystals (rare-gas solids) particularly attractive object for study the low temperature heat capacity and testing theories of lattice dynamics due to the simplicity of cryocrystals. A substantial experimental studies that significantly advanced our understanding of the thermal properties for cryocrystals was obtained in pioneer experimental investigations in previous century and presented in many articles, handbooks and monographies [1 – 7]

A deep understanding of cryocrystals - particularly their lattice dynamics, heat capacity behavior, and quantum vibrational features—is essential for interpreting phenomena in quasi-one-dimensional systems and rotational dynamics in solid matrices. [8-10]. The well-characterized thermal properties of cryocrystals provide a reference and their simplicity make cryocrystals the benchmark for exploring quantum efects and phonon transport in many condenced mater systems [11,12].

The proportional correlation between heat capacity at constant volume $C_v(T)$ and thermal expansion was found for monoatomic cryocrystals (Ar, Kr, Xe) and simple molecular systems (CO, $CO_2$, $N_2O$) attributed to the appearance of the first Van Hove singularity in the phonon density of states. [13]. The normalized heat capacity $C(T)/T^3$ typically exhibits a hump at $T_{max}$ in the range 2–20 K, a feature linked to this singularity [14 – 16]. The position of the hump $T_{max}$ shifted with the change of the molar volume of the Ne cryocrystals due to the change of the phonon density of states and position of the first Van Hove singularity [17].

Behaviour of the low-temperature heat capacity observed for cryocrystals as well as for many other crystalline and amorphous solids [18, 19] has not been fully understood yet. Hump in $C_\mathrm{p}/T^3$ have been reported in a wide variety of crystalline and amorphous materials, often associated with excess low-frequency vibrational states (boson peaks), phonon softening, or enhanced anharmonicity [18 – 22]. Interaction between acoustic and optical phonons considered as the nature of the boson peak for the low-density amorphous LDA ice and hexagonal crystalline ice ($I_h$) [20]. Maximum for heat capacity at constant pressure $C_p$ in the representation $C_p/T^3$ versus $T$ for amorphous solids is a consequence of the excess of low-frequency states corresponding to a local maximum in the reduced vibrational density of states $g(\omega)/\omega^2$ vs $\omega$ [18, 19]. The soft potential model described the hump in $C_p/T^3$ for many amorphous solids. [21]. For large sets of crystalline materials such as superionic conductors, orientationally disordered crystals, ferroelastic memory alloys, metal halide perovskites, ferroelectric materials, organic materials and even molecular crystals without any clear sign of disorder the $C_p/T^3$ hump is rather related to a phonon softening or to strong phonon-phonon interaction due to enhanced anharmonicity [22]. The universal character of these $C_p/T^3$ hump has not yet been conclusively resolved.

Recently, for a set of different materials, including ordered or disordered system and glasses, the universal scaled excess of heat capacity near the hump was demonstrated [23 – 26]. Although universal scaling relation $\Delta^*$ proposed for both disordered and ordered systems [23–25] well describe the hump features of the heat capacity, the microscopic origin and general applicability of this relation remains open question. Correlation between the Debye temperatures and $T_{max}$ temperature was demonstated for some groups of structural glasses and the minimally disordered crystals [23]. M.A. Strzhemechny et al. found the relation between the characteristic temperature of the first van Hove singularity $\Theta_\mathrm{vH}$ and the temperature $T_{max}$ with some empirical ratio $\Theta_\mathrm{vH}/T_\mathrm{max} \approx 5$ observed for classical atomic (Ar, Ne, Kr, and Xe) and molecular ($N_2$, CO, $CO_2$, $N_2O$) cryocrystals [16]. The universal empirical scaled relation $\Delta^*$ was used to describe the non-Debye excess in isochoric heat capacity at low temperatures for sets of Ne cryocrystals with different molar volumes [17]. Such scaled relation $\Delta^*$ is well applied also for other substances [21, 25, 26].

The question about the relation between characteristic temperature of the first van Hove singularity $\Theta_{vH}$, and temperature of the peak $T_{max}$ as well as the question about the relation between the $\Theta_{vH}$ and $\omega_D$ needs further study including considering the quantum cryocrystrals. Addressing this question requires a comprehensive analysis of cryocrystals—encompassing both classical atomic and quantum systems—within the framework of the universal function $\Delta^*$. This approach enables the evaluation of pressure-induced effects on the low-temperature behavior of heat capacity, particularly near the hump observed in the temperature dependence of the normalized heat capacity $C/T^3$.

Motivation for considering the quantum cryocrystals related to features in the heat capacity due to the zero-point energy. Quantum cryocrystals such as solid helium and hydrogen to their extremely light atomic masses present an even greater challenge. Their zero-point energy is comparable to the van der Waals interaction energy, strongly influencing lattice dynamics and shifting the hump in $C(T)/T^3$ with molar volume [27 – 30]. Experimental studies of helium confirm strong anharmonic effects and additional contributions from vacancy excitations [31 – 33]. The relationship between the first Van Hove singularity ($\Theta_{vH}$), the Debye frequency ($\omega_D$), and the temperature of maximum for the hump ($T_{max}$) in such quantum systems remains insufficiently understood.

In this study we extend investigations of the universal empirical scaled relation $\Delta^*$ to set of the atomic cryocryslats and quantum cryocrystals (e.g., $^4$He, $^3$He, $H_2$).. The zero-point energy for quantum cryocrystals (helium, hydrogen) is comparable to the potential energy of intermolecular van der Waals interactions [27]. From the lattice dynamical point of view, zero-point energy fluctuation makes the complication for studying the influence of the change of the magnitude of the first van Hove singularity in the $g(\omega)$ with the change of the molar volume of crystals on the behavior of the heat capacity and can lead to additional features. For solid parahydrogen in the heat capacity at constant volume $C_v(T)/T^3$ [28], the temperature $T_{max}$ for the hump described by power law function:

$$T_{max}(V_1)/T_{max}(V_2)=(V_2/V_1)^{\gamma}, \qquad (1)$$

where $V$ is the molar volume and $\gamma = -\,\partial(\ln\omega_D)/\partial(\ln V)$ is a Grüneisen parameter [13]. It is indicated the shift of the $T_{max}$ to lower temperatures with the increase of the molar volume.

Helium isotopes as well as hydrogen have very light mass of atoms and a weak interaction with the lattice particles [29]. The small mass of these substances and the weakness of the attractive force of their van der Waals interaction compared to the quantum-mechanical zero-point energy lead to the low boiling temperatures and critical points [30]. According to the pressure –temperature phase diagrams for $^4$He at temperature 1 K and pressure above 26 bar, it is solids without mass flow [31]. At temperature 4 K, the pressure is above 125 bar for $^4$He solids, respectively [30 – 33].

The heat capacity of hcp helium has been measured for different volumes at temperatures up to the melting point by Dugdale [34], Edwards [35], and Ahlers [36]. Investigation of the hump in $C/T^3$ of helium is difficult due to zero-point vibrations and the anharmonicities for quantum crystal that affect much stronger than in the heavier rare gas crystals. In addition, the contribution of the vacancy excitations to the heat capacity in solid helium increases with the increase of the molar volume of cryocrystals [37].

The aim of this work is to extend the concept of the universal scaling relation $\Delta^*$ beyond neon to a broader set of cryocrystals, including both classical atomic (Ar, Kr, Xe) and quantum systems ($^4He$, $^3He$, $H_2$). By analyzing the effect of pressure and molar volume on the low-temperature behavior of $C/T^3$, we search the role of Van Hove singularities and zero-point fluctuations in shaping the universal features of heat capacity.

## 2. Atomic Rare-Gas (Cryocrystals)

Rare gases (Ne, Ar, Kr, Xe) form model classical cryocrystals, characterized by weak, isotropic van der Waals interactions between atoms. Their low-temperature heat capacity has been extensively measured over decades, resulting in well-established and consistent recommended values for constant pressure $C_p(T)$ and volume $C_v(T)$ [38 – 51].

The experimental data for these systems are: for Xe, by Clusius [48], Packard [42], Trefny [43], and Fenichel & Serin [38], with later refinements by Klein [1]; for Ar, by Clusius [48], Flubacher et al. [47], Finegold & Phillips [45], and Bagatskii et al. [38]; for Kr, by Clusius [40], Finegold & Phillips [37], and Beaumont et al. [50]; and for Ne, by Fenichel & Serin [28] and Fugate & Swenson [39]. In each case, the recommended $C_v(T)$ curves were derived by smoothing and correcting experimental $C_p(T)$ data for thermal expansion.

At the lowest temperatures, the lattice (phonon) heat capacity of solids is dominated by long-wavelength acoustic phonons with a linear dispersion relation. This leads to the Debye $T^3$ -law:

$$C(T) = \mathrm{C}_{\mathrm{Deb}} T^3, \tag{3}$$

where $\mathrm{C}_{\mathrm{Deb}} = 12\pi^4 R/5\Theta_D^3$, and $R$ is the gas constant and $\Theta_D$ is the Debye temperature.

However, the Debye model is an approximation that assumes a parabolic phonon density of states (DOS), $g(\omega) \propto \omega^2$. In real crystals, the dispersion relation $\omega(k)$ is more complex [52,53]. The actual density of states (DOS) deviates from this simplified form due to the complexities of lattice dynamics, particularly near boundaries of the Brillouin zone where phonon group velocities diminish or vanish, leading to van Hove singularities. Van Hove singularities appear as kinks or peaks in $g(\omega)$.

As temperature increases, these singularities cause systematic deviations from the Debye $T^3$ -law. This is clearly observed as a characteristic hump in a plot of $C(T)/T^3$ versus $T$, with a maximum at a temperature $T_{max}$ [14 – 16]. This deviation is common in both atomic rare-gas cryocrystals and molecular cryocrystals like $N_2$ and $CO_2$.

The phonon heat capacity at low temperatures can be more accurately described by an expansion series:

$$g(\omega)=\alpha\omega^2+b\omega^4+\dots \quad (5)$$

$$C_{ph} = C_{Deb}T^3+C_5T^5+\dots \quad (6)$$

The higher-order term $C_5$ arises from the non-linearity of the acoustic phonon dispersion relation near the zone edges [14 – 16]:

$$\omega^2(q) = s^2q^2 - k^2q^4, \quad (7)$$

with $s$ denoting the sound velocity and $k$ the rigidity parameter.

Thus, while rare-gas cryocrystals are nearly ideal van der Waals solids, their heat capacities exhibit deviations from the simple Debye model. This complex behavior provides the essential baseline for understanding the even more complicated quantum effects observed in cryocrystals like hydrogen and helium.

## 3 Quantum cryocrystals

Both parahydrogen and solid helium are quantum cryocrystals, where light atomic/molecular mass, weak bonding and large zero-point energy fundamentally affect the low temperature behavior of heat capacity. Nevertheless, despite their quantum character, they share with classical rare-gas solids the common feature of a $C_v/T^3$ hump, underscoring the universality of vibrational thermodynamics across cryocrystals.

### 3.1. Parahydrogen (p-$H_2$).

The specific heat of solid parahydrogen has been studied in detail by Roberts and Daunt [54] and by Bagatskii et al. [55] in the range 0.5–8 K, and more extensively by Krause and Swenson [28] over 4 K to the melting line for molar volumes between 22.79 and 16.19 cm$^3$/mol. At the lowest temperatures, the heat capacity follows the Debye $T^3$-law, while deviations appear at higher $T$ due to phonon dispersion and anharmonicity. Compared with heavier rare-gas solids, the deviations set in earlier due to the lighter molecular mass and lower Debye temperature of hydrogen. Moreover, orientational degrees of freedom and weak rotational tunneling make p-$H_2$ an intermediate case between purely atomic cryocrystals and molecular quantum solids [28, 54, 55].

### 3.2. Solid helium isotopes.

Solid helium illustrates the most pronounced manifestation of the quantum lattice behavior, forming crystal only under applied pressures exceeding approximately 25 atm. This behavior arises from pronounced quantum-zero-point vibrations, which counteract the weak van der Waals interactions and inhibit crystallization under ambient conditions. Estimates suggest that in solid $^4$He, nearly 98% of the ground-state energy arises from zero-point vibrations, with only ~2% from interatomic binding. This places helium at the ultimate quantum limit, where classical solid-state concepts are inadequate.

The heat capacity of helium isotopes was first measured by Dugdale and Franck [34], with high-precision $^4$He data later obtained by Edwards and Pandorf [35] and Ahlers [36]. At very low temperatures, both $^3$He and $^4$He display the cubic Debye law, but the validity range is restricted to only a few kelvin because of their very low Debye temperatures ($\Theta_D$~25–30 K for $^4$He near melting). At higher $T$, $C_v(T)$ departs much more strongly from the Debye prediction than in heavier cryocrystals, producing a pronounced maximum in $C_v/T^3$. The isotope effect for helium appears due to the lighter $^3$He exhibiting lower vibrational frequencies and larger heat capacity at the same temperature, reflecting the enhanced role of zero-point motion.

Beyond phonons, helium shows an additional anomaly at temperatures near melting, first recognized by Simon [56] and later elaborated by Grigor'ev and co-workers [37]. This contribution originates from thermal vacancies, whose equilibrium concentration is anomalously large because the vacancy formation energy is only 10 – 12 K in $^4$He (in contrast to hundreds and thousands Kelvin in classical solids).

Thus, the heat capacity of solid helium provides a unique object for quantum physics. It reflects the interplay of zero-point motion, phonon dispersion, anharmonicity, isotope effects, and vacancy excitations. In contrast to rare-gas cryocrystals, where vacancy effects are negligible and the Debye behavior of heat capacity appears at low temperatures, helium demonstrates the essential role of quantum fluctuations and defects in shaping its thermal properties.

## 4. Analysis of Low-Temperature Heat Capacity Data

Let us turn to a comparative analysis of the low-temperature heat capacity behavior of classical rare-gas cryocrystals and quantum cryocrystals, building on the experimental overview in Section 2 for classical rare-gas cryocrystals and Section 3 for quantum cryocrystals.

The heat capacity $C_v$ of all analyzed cryocrystals (Xe, Ar, Kr, Ne, $H_2$, $^3$He, and $^4$He) can be described by an expansion series (eq. 4) with the Debye term, proportional to $T^3$, and additional contributions that manifest as characteristic humps in the $C_v/T^3$ versus $T$. The hump $C_v/T$ is clearly displayed in Fig. 1. The inset in Fig. 1 illustrates the contrast between the idealized phonon frequency distribution of the Debye model, $g_D(\nu)/\nu^2$, and the actual vibrational spectrum $g(\nu)/\nu^2$ obtained for solid Kr. The experimental spectrum $g(\nu)$ for solid Kr, determined by Skalyo Jr. et al. through inelastic neutron scattering at 10 K, reveals the first Van Hove singularity in $g_D(\nu)/\nu^2$ at $\nu$=0.7 THz [57]. The characteristic temperature corresponding to this singularity is $\Theta_{vH}\approx 33.6$ K. The hump in $C_v/T^3$ for Kr is observed at $T_{max}$=6.25 K, yielding a ratio $\Theta_{vH}/T_{\max}\approx5$. This scaling, with $\Theta_{vH}/T_{max}\approx5$, is a common feature not only for classical atomic cryocrystals (Ar, Ne, Kr, Xe) but also for molecular cryocrystals such as $N_2$, CO, $CO_2$, and $N_2O$ [16].

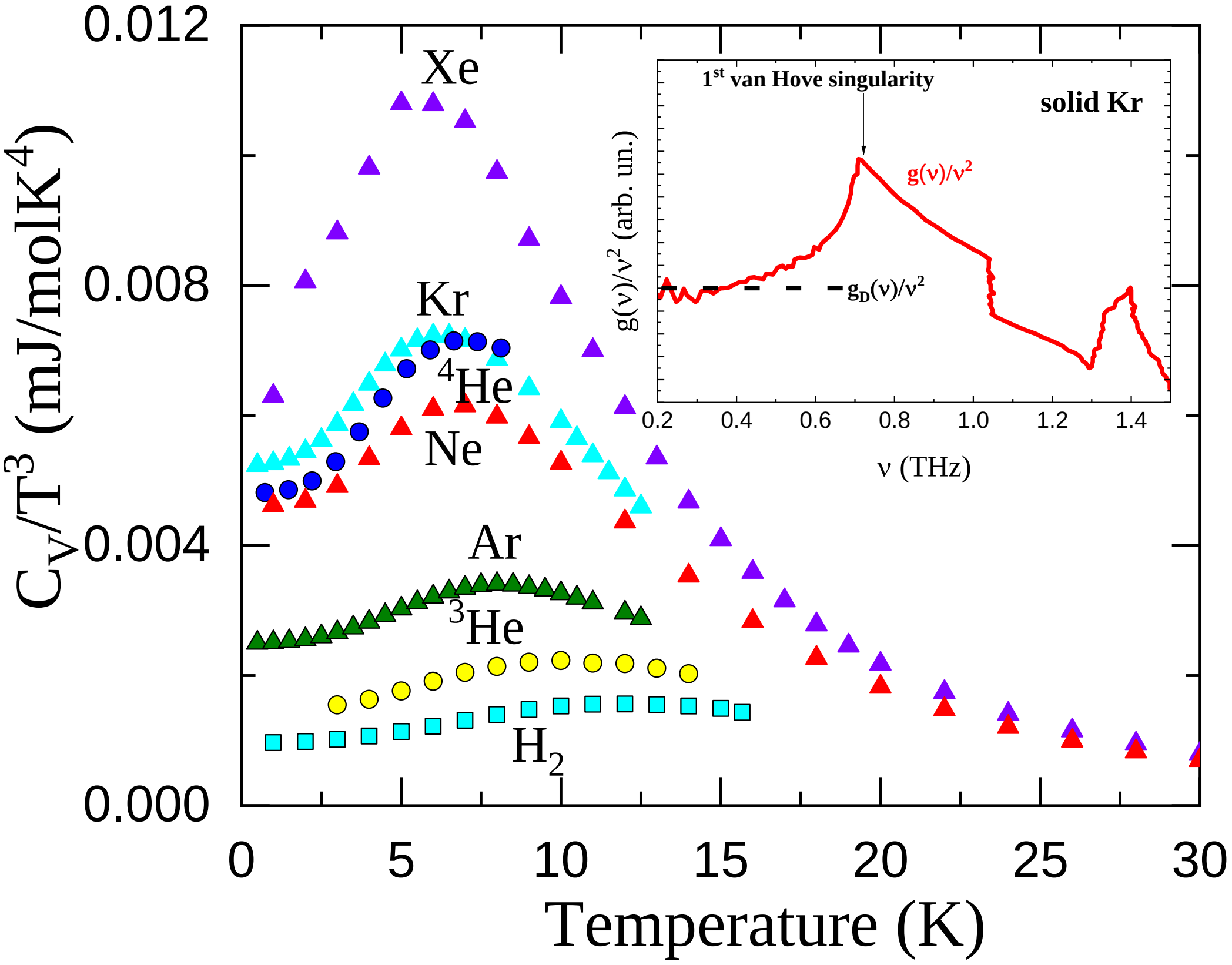

Fig. 1. Normalized heat capacity $C_V/T^3$ vs. $T$ for: ▲ – Xe (V = 34.73 cm$^3$/mol) [44], ▲ – Kr (V = 26.932 cm$^3$/mol) [45], ● – $^4$He (V = 13.723 cm$^3$/mol) [36], ▲ – Ne (V = 13.39 cm$^3$/mol) [39], ▲ – Ar (V = 22.415 cm$^3$/mol) [45], ● – $^3$He (V = 12.57 cm$^3$/mol) [34], ■ – $H_2$ (V = 22.787 cm$^3$/mol) [28]. Insert: The red straight line shows the sketch of the calculated real spectra with Van Hove singularities in the density of vibrational states $g(\nu)/\nu^2$ vs. $\nu$ for solid Kr [57]; the black dot line shows the sketch of the phonon frequency distribution $g_D(\nu)/\nu^2$ vs. $\nu$ for the Debye model.

Experimental values $T_{max}$, $C_{Deb}$, $[C/T^3]_{\max}$, the fitting parameter D=3 [17], and the parameter $d=C_{Deb}/[C/T^3]_{\max}$ for the investigated cryocrystals are summarized in Table 1. It is seen that both $[C/T^3]_{\max}$ and $C_{Deb}$ decrease systematically with increasing $T_{max}$. The dependence of $d=C_{\rm Deb}/[C/T^3]_{\max}$ vs. $T_{max}$ is plotted in Fig. 2, where the values cluster around constant $d_{eff}\approx0.69$. The largest deviations from this universal value observed for cryocrystals with lightest mass of particles: for $H_2$, the value of $d$ is approximately 11% lower than $d_{eff}$, while for Ne it is about 7% higher. The observed uniformity of the scaling parameter $d$ across diverse cryocrystals highlights the validity and resilience of the scaling law. Conversely, deviations from $d_{eff}$ serve as a sensitive diagnostic of quantum mechanical contributions to lattice dynamics.

This analysis therefore unifies the behavior of both classical rare-gas cryocrystals (Section 2) and quantum cryocrystals (Section 3): despite differences in binding strength, zero-point motion, and anharmonicity, all systems exhibit the universal $C_v/T^3$ hump governed by phonon dispersion and Van Hove singularities. The near-constant scaling parameter $d$ demonstrates that the low-temperature heat capacity of cryocrystals is controlled by a common vibrational mechanism, with quantum effects introducing only quantitative, rather than qualitative, modifications.

Table 1. $T_{max}$, $C_{Deb}$ and $[C/T^3]_{max}$, $d=C_{Deb}/[C/T^3]_{max}$, D for cryocrystals.

| Material | $T_{max}$ | $C_{Deb}$, mJ/molK4 | $[C/T^3]_{max}$, mJ/molK4 | $d=C_{Deb}/[C/T^3]_{max}$ | D | Ref. |
|---|---|---|---|---|---|---|
| Xe (cubic, fcc, 6.132Å) [1] V=34.73 $cm^3$/mol | 5.83 | 7.41 | 10.8 | 0.68 | 3 | [44] |
| Kr (cubic, fcc, 5.646Å) [1] V=26.932 $cm^3$/mol | 6.25 | 5.23 | 7.24 | 0.72 | 3 | [45] |
| $^4$He hcp, V=13.727 $cm^3$/mol | 6.93 | 4.91 | 7.15 | 0.67 | 3 | [36] |
| Ne hcp, V=13.39 $cm^3$/mol | 6.72 | 4.61 | 6.17 | 0.75 | 3 | [39] |
| Ar (cubic, fcc, 5.311Å) [1] V=22.415 $cm^3$/mol | 8.05 | 2.5 | 3.41 | 0.73 | 3 | [45] |
| $^3$He hcp, V=12.57 $cm^3$/mol | 10 | 1.53 | 2.23 | 0.67 | 3 | [34] |
| $H_2$ (V = 22.787 $cm^3$/mol) | 11.66 | 0.97 | 1.56 | 0.62 | 3 | [28] |

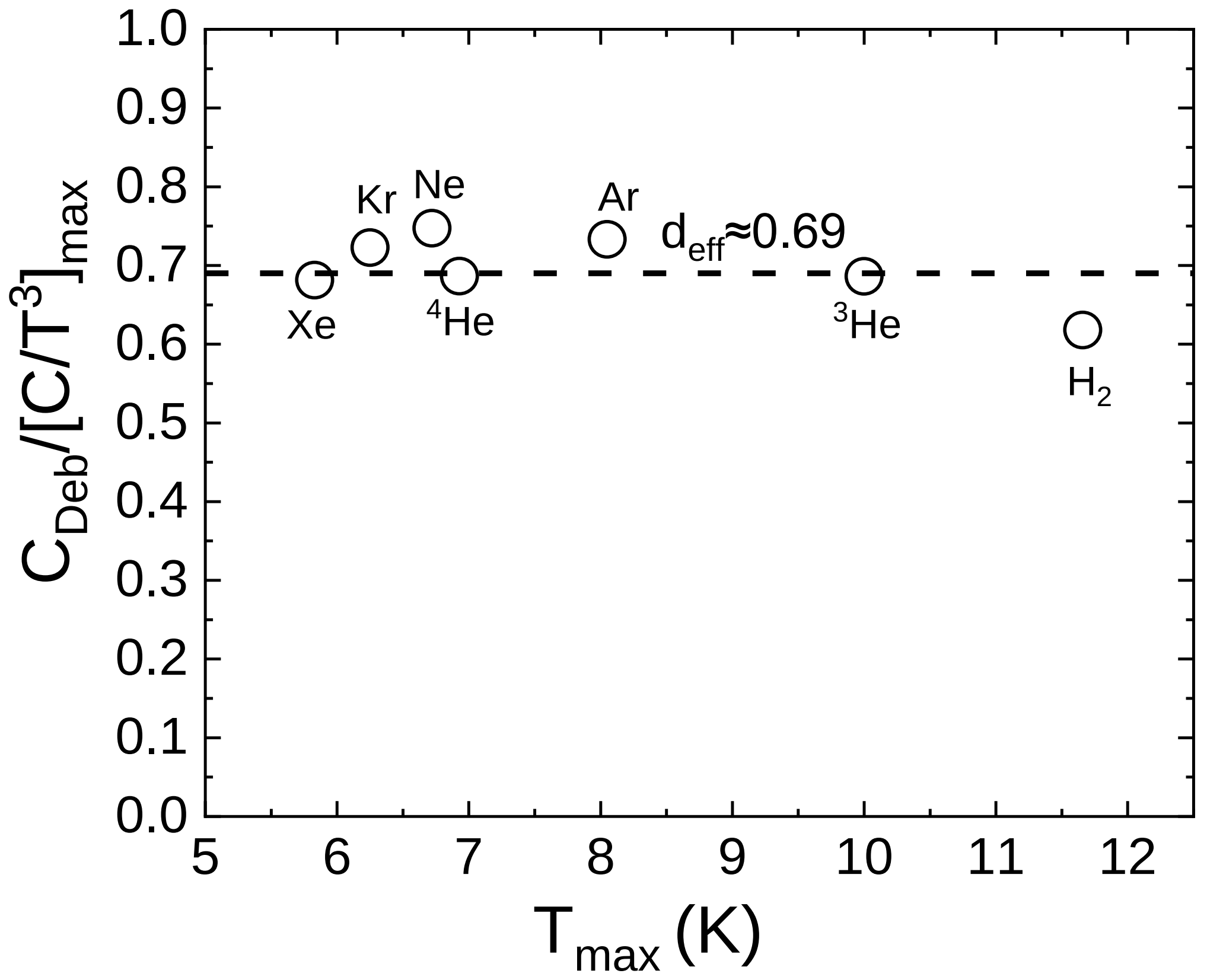

Fig. 2 $C_{Deb}/[C/T^3]_{max}$ vs. $T_{max}$ for cryocrystals Xe, Ar, Kr, Ne, $H_2$, $^3$He and $^4$He.

Fig. 3 demonstrates that both $C_{Deb}$ and $[C/T^3]_{max}$ decrease approximately linearly in log-log scale with increasing $T_{max}$. This proportional decrease, coupled with the nearly constant parameter $d$, implies that the characteristic temperatures $\Theta_D$, $\Theta_{vH}$, and $T_{max}$ scale consistently with changes in molar volume across all cryocrystals:

$$\frac{\Theta_D(V_1)}{\Theta_D(V_2)} = \frac{\Theta_{\mathrm{vH}}(V_1)}{\Theta_{\mathrm{vH}}(V_1)} = \frac{\mathrm{T_{max}}(V_1)}{\mathrm{T_{max}}(V_1)}, \tag{8}$$

A similar linear correlation between the Debye temperature and the hump position in $C/T^3$ was predicted by Granato for structural glasses and minimally disordered crystals [23, 58, 59].

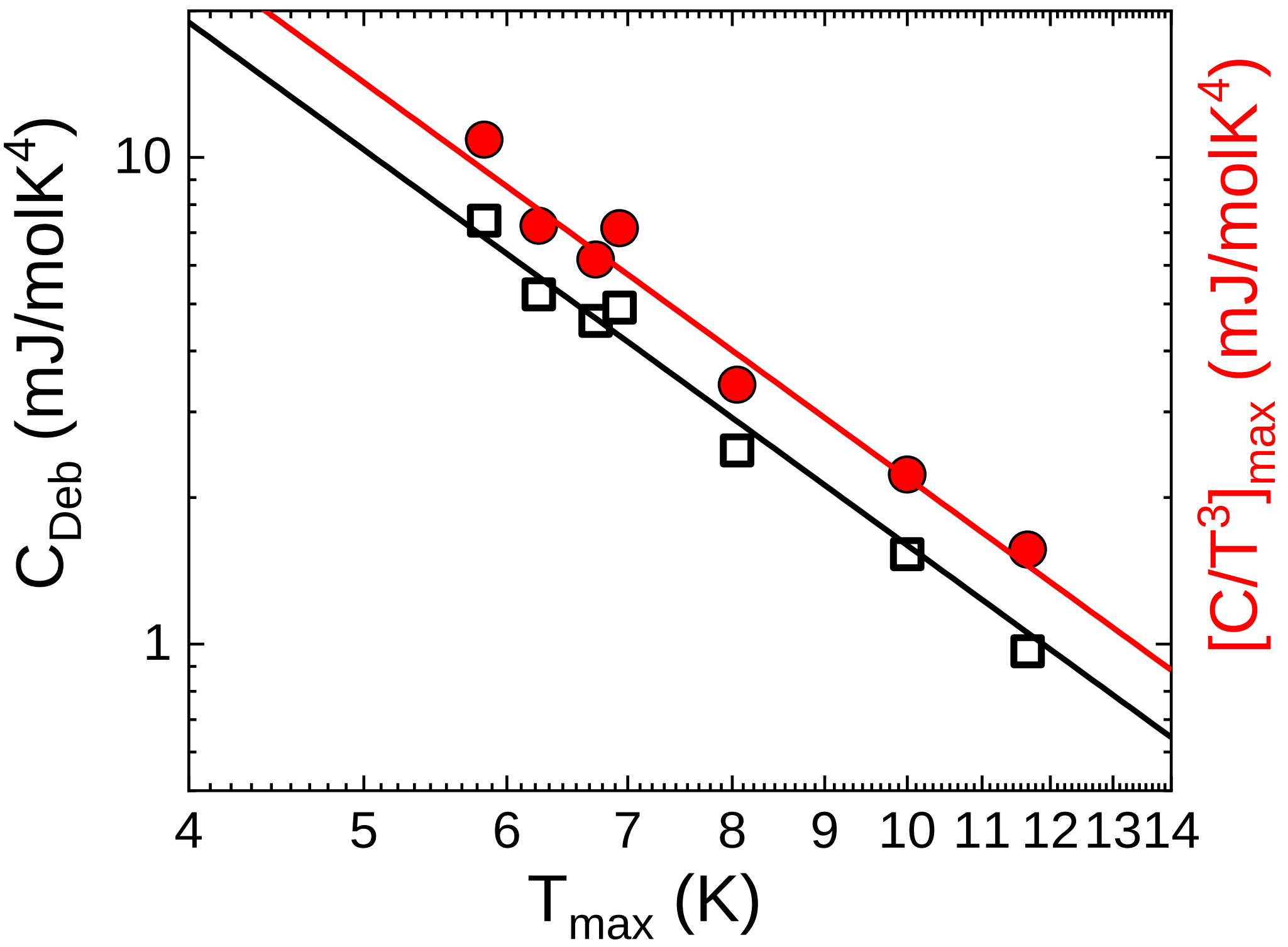

Fig. 3 □ - $C_{Deb}$, ● - $[C/T^3]_{max}$ vs. $T_{max}$ for cryocrystals Xe, Ar, Kr, Ne, $H_2$, $^3$He and $^4$He based on the data in the table 1.

The universality of this behavior can be probed more quantitatively by analyzing the normalized ratio $[C/T^3]/[C/T^3]_{max}$ as a function of $T/T_{max}$. As shown in Fig. 4 (a),(b), the behavior of normalized ratio for all cryocrystals described a common quadratic dependence $y=1-d(1-T/T_{max})^2$ in the vicinity of the hump ($0.5<T/T_{max}<1.5$). Deviations occur at temperatures $T/T_{max}<0.2$, that reflecting differences in the Debye temperature of each system. For simple atomic cryocrystals with fcc structure (Ar, Kr, Xe), a systematic reduction in $d=C_{Deb}/[C/T^3]_{max}$ is observed with increasing lattice parameter, which is consistent with the shrinking of Brillouin zone.

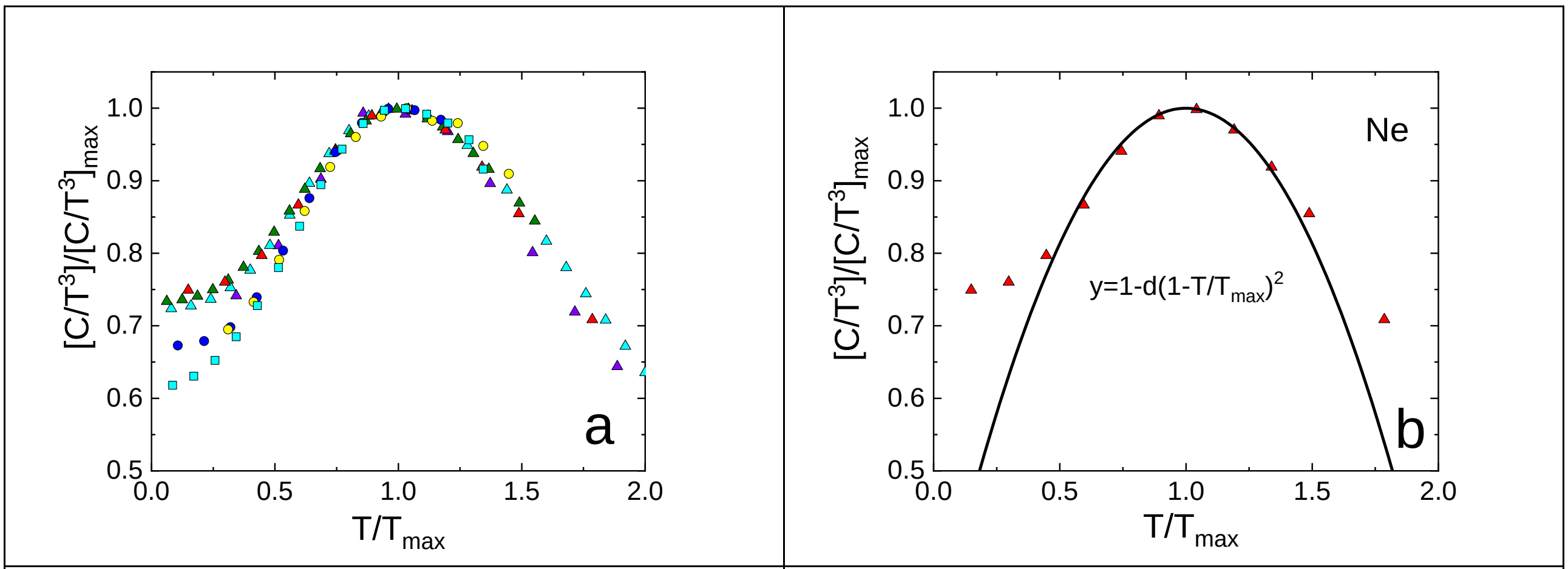

| a | b |
|---|---|

Fig. 4. a) Ratio $[C/T^3]/[C/T^3]_{max}$ vs. $T/T_{max}$: for: ▲ – Xe ($V = 34.73$ cm$^3$/mol [44]; ▲ – Kr ($V = 26.932$ cm$^3$/mol [45]; ● – $^4$He ($V = 13.723$ cm$^3$/mol) [36]; ▲ – Ne (V = 13.39 cm$^3$/mol) [39]; ▲ – Ar ($V = 22.415$ cm$^3$/mol [45]; ● – $^3$He (V = 12.57 cm$^3$/mol) [34]; ■ – $H_2$ ($V = 22.787$ cm$^3$/mol) [28]; b) Ratio $[C/T^3]/[C/T^3]_{max}$ vs. $T/T_{max}$ для: ▲ – Ne ($V = 13.39$ cm$^3$/mol) [39]. Function $y=1-d(1-T/T_{max})^2$ symmetrically describes the hump in such coordinates for coefficient $d = C_{Deb}/[C/T^3]_{max}$ that is taken from ratio $[C/T^3]/[C/T^3]_{max}$ vs. $T/T_{max}$ at $T/T_{max}\to 0$.

A more general universal description is provided by the dimensionless function $\Delta^*(T/T_{max}, C(T), C_{Deb}, C(T_{max}))$ defined as [25]:

$$\Delta^*\left(\frac{T}{T_{max}}, C(T), C_{Deb}, C(T_{max})\right) = \frac{T_{max}^3}{T^3}\frac{C(T) - C_{Deb}\,T^3}{C(T_{max}) - C_{Deb}\,T_{max}^3}\quad , \qquad (7).$$

As shown in Fig. 5, $\Delta^*$ behaves universally, approaching zero as $T/T_{max}\to 0$ and equal one as $T/T_{max}\to 1$. Near the hump, all cryocrystals exhibit a symmetric quadratic dependence $\Delta^*=1-D(1-T/T_{max})^2$ with a universal fitting parameter D=3.

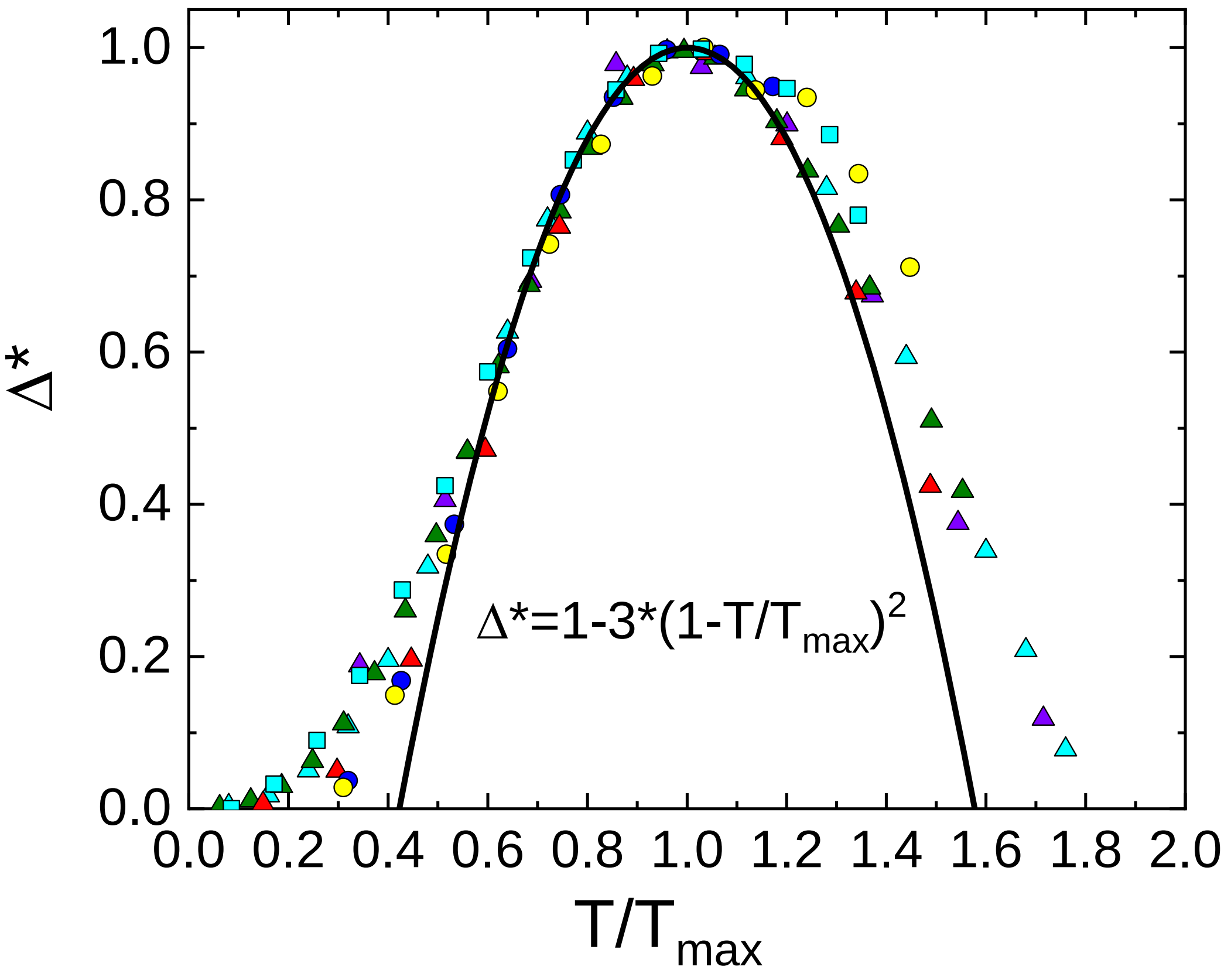

Fig 5. $\Delta^*$ vs. $T/T_{max}$ for: ▲ – Xe (V = 34.73 cm$^3$/mol [44]; ▲ – Kr (V = 26.932 cm$^3$/mol [45]; ● – $^4$He (V = 13.727 cm$^3$/mol) [36]; ▲ – Ne (V = 13.39 cm$^3$/mol) [39]; ▲ – Ar (V = 22.415 cm$^3$/mol [45]; ● – $^3$He (V = 12.57 cm$^3$/mol) [34]; ■ – $H_2$ (V = 22.787 cm$^3$/mol) [28];

Additional evidence for universality is provided in Fig. 6, which plots $\Delta^*$ vs. $(1-T/T_{\rm max})^2$ for Xe [44], Kr [45], Ne [39], Ar [45], $^3$He [34], $^4$He [36], and $H_2$ [28]. In the regime $(1-T/T_{max})^2<0.1$, all systems in Fig. 6 follow a linear dependence describing $\Delta^*=1-D(1-T/T_{\rm max})^2$ with D=3. The fact that both the scaling parameter $d\sim d_{\rm eff}$ and the constant D~3 retain nearly constant values both for classical and quantum cryocrystals suggests that shifts in the van Hove singularity due to pressure-induced changes in the Brillouin zone act uniformly on the vibrational density of states $g_D(\omega)=\alpha\omega^2$, with pressure modifying the prefactor $\alpha$. Indeed, the pressure dependence of the hump parameters in Ne has recently been confirmed in [17].

Finally, small asymmetries in the low- and high-temperature sides of the hump are observed for helium and hydrogen. These deviations can be traced to quantum effects, including large zero-point energies and vacancy contributions [27, 37], which become significant due to the unusually low melting and triple points of these systems.

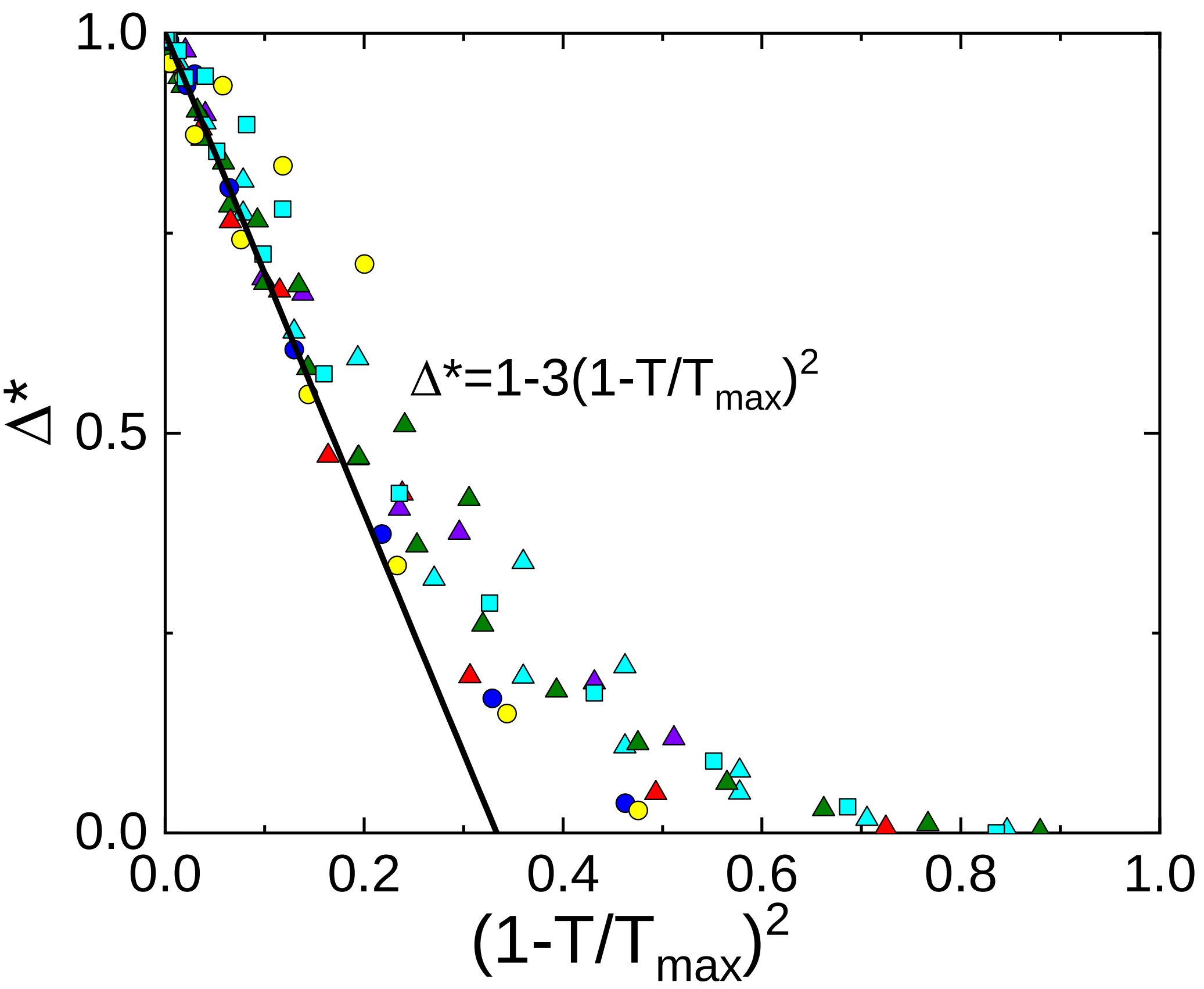

Fig 6. Δ* vs. $(1-T/T_{max})^2$ for: ▲ – Xe (V = 34.73 $cm^3$/mol [60]) [44]; ▲ – Kr (V = 26.932 $cm^3$/mol [60]) [45]; ● – $^4$He (V = 13.727 $cm^3$/mol) [36]; ▲ – Ne (V = 13.39 $cm^3$/mol) [39]; ▲ – Ar (V = 22.415 $cm^3$/mol [60]) [45]; ● – $^3$He (V = 12.57 $cm^3$/mol) [34]; ■ – $H_2$ (V = 22.787 $cm^3$/mol) [28].

The comparative analysis of low-temperature heat capacity data across rare-gas and quantum cryocrystals reveals a striking universality. All systems exhibit a characteristic hump $C/T^3$ linked to the first van Hove singularity, with the position $T_{max}$ and amplitude governed by scaling relations involving $\Theta_D$ and $\Theta_{vH}$. The normalized function $\Delta^*$ further emphasizes this universality, collapsing diverse data sets onto a single quadratic form. Normalization procedures ($[C/T^3]/[C/T^3]_{max}$, $\Delta^*$) collapse diverse data sets onto universal quadratic forms, parameterized by nearly constant coefficients $d_{eff}$ and D=3. Deviations arise only in quantum solids such as helium and hydrogen, where vacancy and zero-point effects contribute. Thus, despite differences in crystal structure, lattice parameter, and quantum character, the fundamental scaling relations provide a unified framework for describing low-temperature heat capacity in cryocrystals, which will be further connected to broader classes of crystalline and glassy solids.

### 4.1. Density effect on the hump in the heat capacity of the $^3$He

Dugdale et al. [34] measured the heat capacity of solid $^3$He for a wide range of molar volumes (V = 12.57, 13.33, 13.56, 14.11, 14.16, 14.98, and 15.72 cm$^3$/mol) at temperatures above 3 K. Since measurements were limited to relatively high temperatures, this leads for significant uncertainty in the direct determination of the Debye coefficient $C_{\mathrm{Deb}}$. As is well known, the relation $C=C_{Deb}T^3$ holds reliably only at $T<\Theta_D/50$. For body-centered cubic (bcc) $^3$He with V=20.18cm$^3$/mol, the Debye temperature is theoretically estimated as $\Theta_D\approx 30$K [61].

This limitation is important in cases where the temperature interval of the experiment is insufficient for the precise determination of both parameters $C_{Deb}$ and $[C/T^3]_{\mathrm{max}}$ directly from the heat capacity data. The $[C/T^3]_{\mathrm{max}}$ is extracted from the maximum of the $C/T^3$ vs. $T$ plot, whereas $C_{Deb}$ is obtained as the slope of the $C/T$ vs. $T^2$ plot. Gardner et al. [62] measured the heat capacity of solid $^4$He between 0.35 – 2 K for several molar volumes (V = 20.96, 20.742, 20.521, and 20.456 cm$^3$/mol). In this case, the temperature range was sufficient to determine $C_{Deb}$ accurately from $C/T$ vs. $T^2$, but too narrow to find the $[C/T^3]_{\mathrm{max}}$.

Before discussing $^4$He, we first analyze $^3$He due to the close physical properties of these two isotopes. Figure 7 shows the experimental $C/T^3$ vs. $T$ data for $^3$He at different molar volumes [34]. The results clearly demonstrate that the maximum position $T_{max}$ decreases with increasing molar volume, in the same way as previously observed for Ne [17]. At elevated molar volumes (V = 14.98 and 15.72 cm$^3$/mol), quantum effects arising from zero-point energy hinder the detection of a clear $T_{max}$. For denser samples (V = 12.57, 13.33, 13.56, 14.11, and 14.16 cm$^3$/mol), the hump in $C/T^3$ vs. $T$ is clearly observed, and $T_{max}$ can be identified directly from experiment. However, since all measurements were performed above 3 K, significant uncertainties remain in the determination of $C_{Deb}$ from the $C/T$ vs. $T^2$ slope. The values of $C_{\mathrm{Deb,exp}}$ obtained this way are summarized in Table 2 and sufficiently agree with theoretical values $C_{\mathrm{Deb,calc}}$ estimated using Eckert's equation for the volume and temperature dependence of the Debye temperature [17, 63].

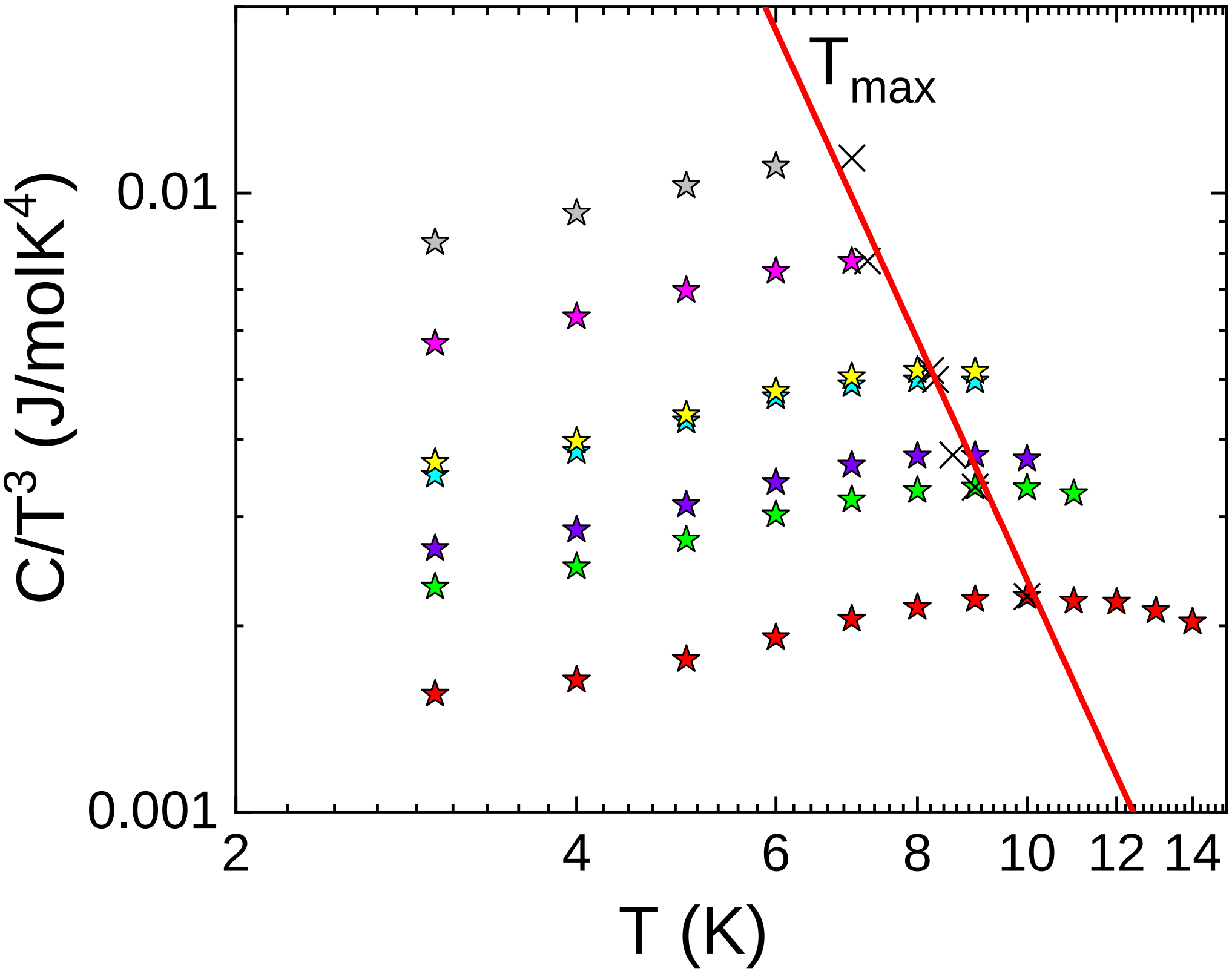

Fig. 7. Normalized heat capacity $C/T^3$ vs. $T$: ★ – $^3$He (V=12.57 cm$^3$/mol) [34]; ★ – $^3$He V=13.33 cm$^3$/mol [34]; ★ – $^3$He (V=13.56 cm$^3$/mol), [34]; ★ – $^3$He (V=14.11 cm$^3$/mol) [34]; ★ – $^3$He (V=14.16 cm$^3$/mol) [34]; ★ – $^3$He (V=14.98 cm$^3$/mol) [34]; ☆ – $^3$He (V=15.72 cm$^3$/mol) [34]. x- Estimated values of $T_{max}$ for $^3$He and line for the eye demonstrates the shift of the $T_{max}$.

Table 2. $T_{max}$, $C_{Deb}$ and $[C/T^3]_{max}$, $d=C_{Deb}/[C/T^3]_{max}$, for $^3$He. *- Estimated values for $T_{max}$, for $^3$He from the shift of the $T_{max}$ with change of the molar volume.

| Material | $T_{max}$ | $C_{Deb,exp}$ mJ/molK$^4$ | $[C/T^3]_{max}$, mJ/molK$^4$ | $d= C_{Deb,exp}/[C/T^3]_{max}$ |
|---|---|---|---|---|
| ★ – $^3$He V=12.57cm3/mol, [34] | 10 | 1.53 | 2.23 | 0.69 |
| ★ – $^3$He V=13.33cm3/mol, [34] | 9.22 | 2.3 | 3.35 | 0.67 |
| ★ – $^3$He V=13.56cm3/mol, [34] | 8.6 | 2.5 | 3.78 | 0.66 |
| ★ – $^3$He  V=14.11cm3/mol, [34] | 8.3 | 3.3 | 5 | 0.66 |
| ★ – $^3$He  V=14.16cm3/mol, [34] | 8.22 | 3.5 | 5.17 | 0.68 |
| ★ – $^3$He V=14.98cm3/mol, [34] | 7.23* | 5.32 | 7.77 | 0.69 |
| ☆ – $^3$He V=15.72cm3/mol, [34] | 7* | 7.8 | 11.4 | 0.68 |

The experimental $C_{\mathrm{Deb,exp}}$ for the sample with V=14.11 cm$^3$/mol was used as a reference to predict $C_{\mathrm{Deb,calc}}$ for the other molar volumes. The maxima $[C/T^3]_{\mathrm{max}}$ are marked by “x” in Fig. 7.

The behavior of $T_{max}$ as a function of molar volume is shown in Fig. 8. For comparison, in solid para-$H_2$ the variation of $T_{max}$ with volume can be described in terms of the Grüneisen parameter $\gamma$ [13]. In Ne, $\gamma$=2.58 is consistent with the data of Fugate and Swenson ($\gamma$=2.51±0.03) [39] and Eckert ($\gamma$=2.60±0.03) [63]. For $^4$He, $\gamma$ ranges from 1.48 to 2.57 depending on molar volume [64]. In the present analysis, we find that $\gamma\approx1.95$ offers a quantitatively consistent representation of the observed $T_{max}(V)$ data for $^3$He. Similarly, for $^4$He, $\gamma\approx2.0$–2.1 reproduces $T_{max}(V)$ from heat capacity and neutron scattering data, and agrees well with Raman measurements [65] and neutron inelastic scattering results [66, 67].

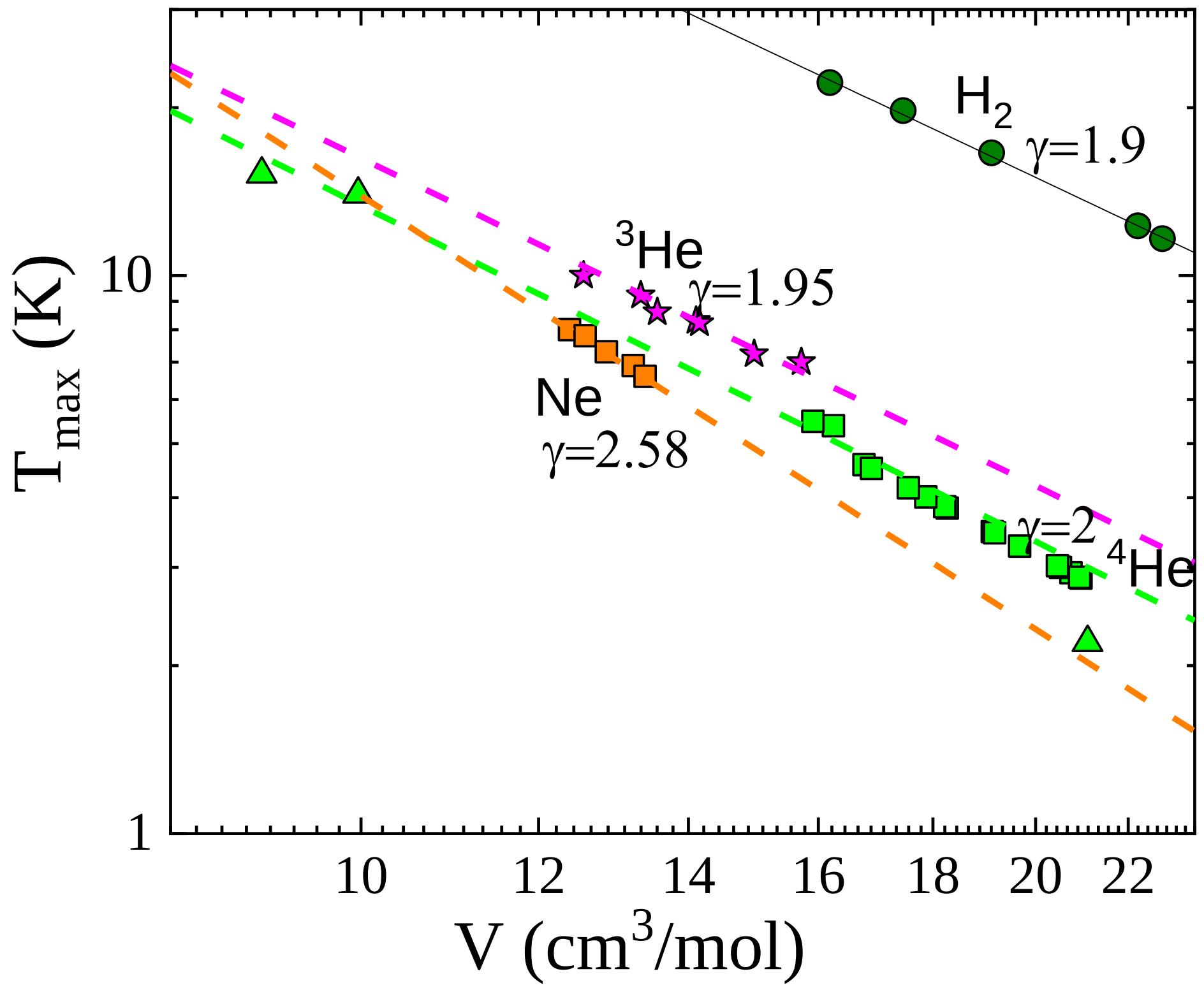

Fig 8. $T_{max}$ vs. $V$ for ■-Ne [39], ★ -$^3$He [34], ■ -$^4$He [34, 35, 36, 62], ● -$H_2$ [28].

The correlation between $C_{Deb}$, $[C/T^3]_{\max}$, and $C_{\mathrm{Deb,calc}}$ with $T_{\max}$ is illustrated in Fig. 9. For $^3$He, these parameters scale as:

$$C_{Deb,V2}/C_{Deb,V1} = (T_{max,V1}/T_{max,V2})^{4.1}, \quad (12)$$

$$[C/T^3]_{\mathrm{max,\ V2}}/[C/T^3]_{\mathrm{max,\ V1}} = (T_{max,V1}/T_{max,V2})^{4.1}, \quad (13)$$

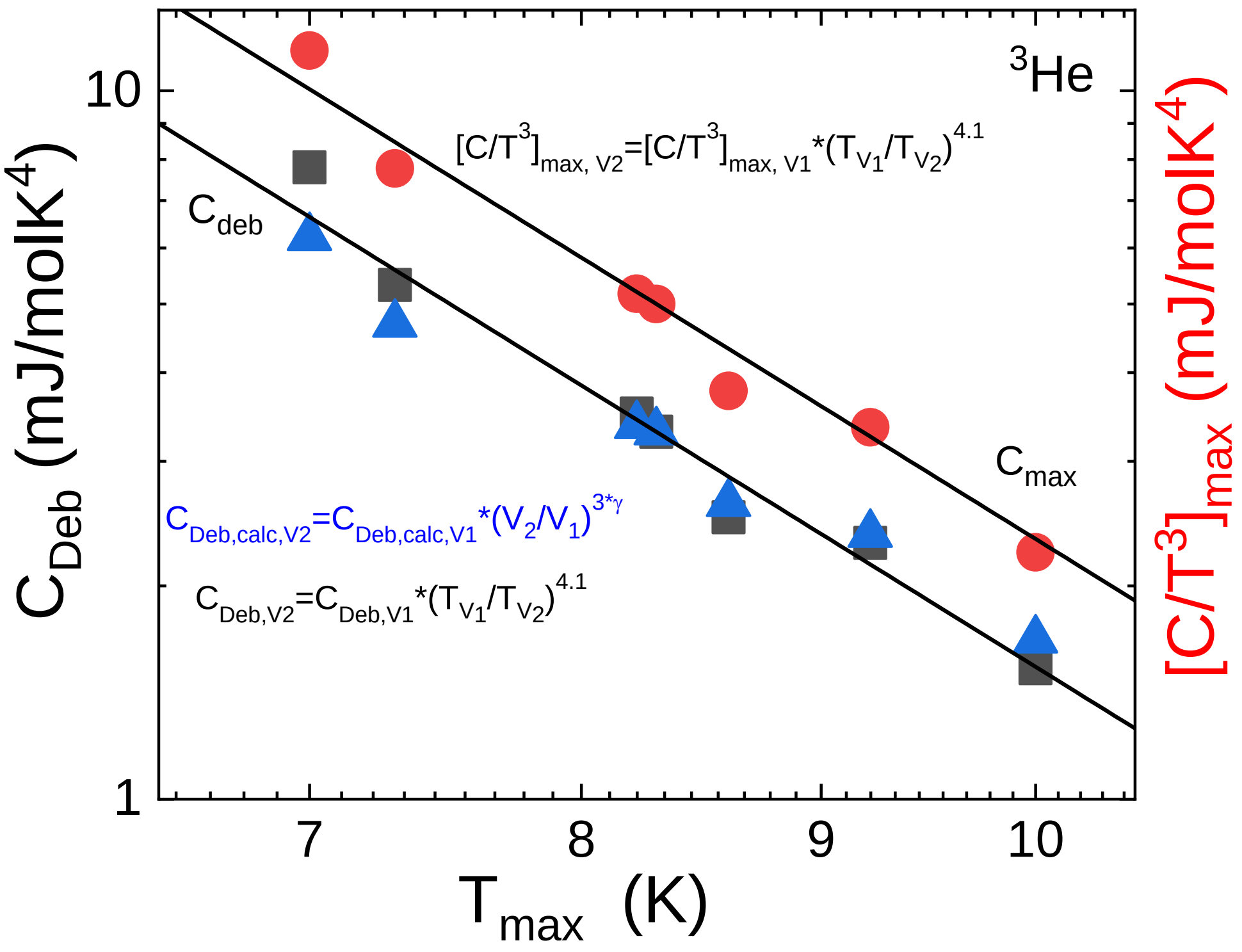

Fig. 9. $C_{Deb}$ , $C_{Deb,calc}$, $[C/T^3]_{max}$ vs. $T_{max}$ for ★ – $^3$He (V = 12.57 cm$^3$/mol) [34]; ☆ – $^3$He V = 13.33 cm$^3$/mol, [34]; ☆ – $^3$He (V = 13.56 cm$^3$/mol), [34]; ☆ – $^3$He (V = 14.11 cm$^3$/mol), [34]; ☆ – $^3$He (V = 14.16 cm$^3$/mol), [34]; ☆ – $^3$He (V = 14.98 cm$^3$/mol), [34]; ☆ – $^3$He (V = 15.72 cm$^3$/mol), [34].

This scaling is consistent with Eckert's model:

$$C_{Deb}(V_2)/C_{Deb}(V_1) \sim (V_2/V_1)^{3\gamma} , \quad (14)$$

using γ=1.95. However, the parameter $d=C_{Deb}/[C/T^3]_{max}$ shows stronger sensitivity for $^3$He than in Ne. While experimental values of $d$ remain nearly constant at 0.69, we predicted that it should systematically decrease from 0.75 to 0.55 with an increase in the molar volume according to Eckert's equation [17, 63]. The discrepancy between experimental $d$ and its theoretical prediction (see fig. 10) likely demonstrates strong quantum effects: the zero-point energy in $^3$He is comparable to the van der Waals potential energy, leading to enhanced fluctuations at larger molar volumes.

Thus, while classical models predict a systematic decrease of $d$ with molar volume, the experimental constancy of $d$ in $^3$He reveals the dominant role of zero-point fluctuations in shaping its thermodynamic behavior. Figure 11 illustrates this by showing the normalized dependence $[C/T^3]/[C/T^3]$max vs. $T/T_{max}$, emphasizing the anomalous robustness of the experimental ratio in $^3$He compared to classical predictions.

The ratio $[C/T^3]/[C/T^3]_{max}$ vs. $T/T_{max}$ for $^3$He is shown in Fig. 11. This representation allows direct comparison of different molar volumes and highlights universal scaling features.

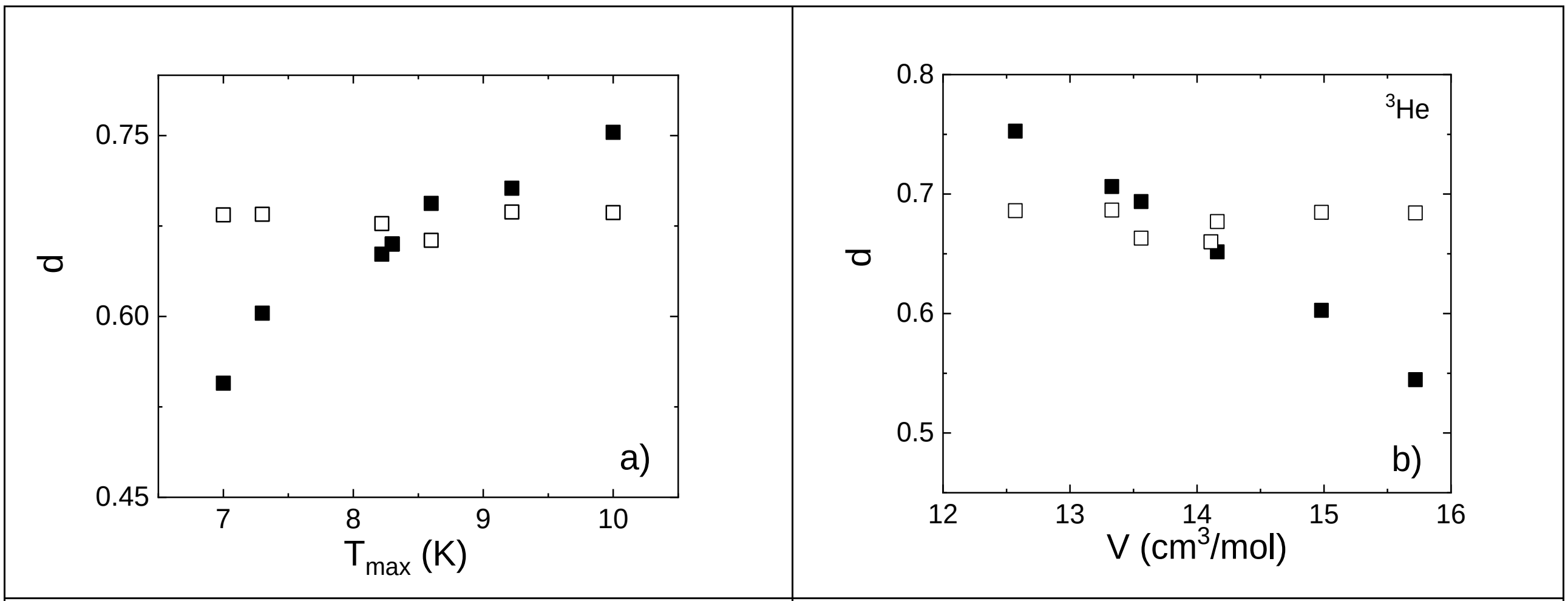

a b

Fig. 10. $^3$He: Experimental □ $d=[C/T^3]/[C/T^3]_{\max}$ and calculated ■ $d_{calc}=C_{Deb,calc/}/[C/T^3]_{\max}$ ratio vs. $T_{max}$ (a) and Experimental $d=[C/T^3]/[C/T^3]_{\max}$ and calculated $d_{calc}=C_{Deb,calc}/[C/T^3]_{\max}$ ratio vs. $V$ (b).

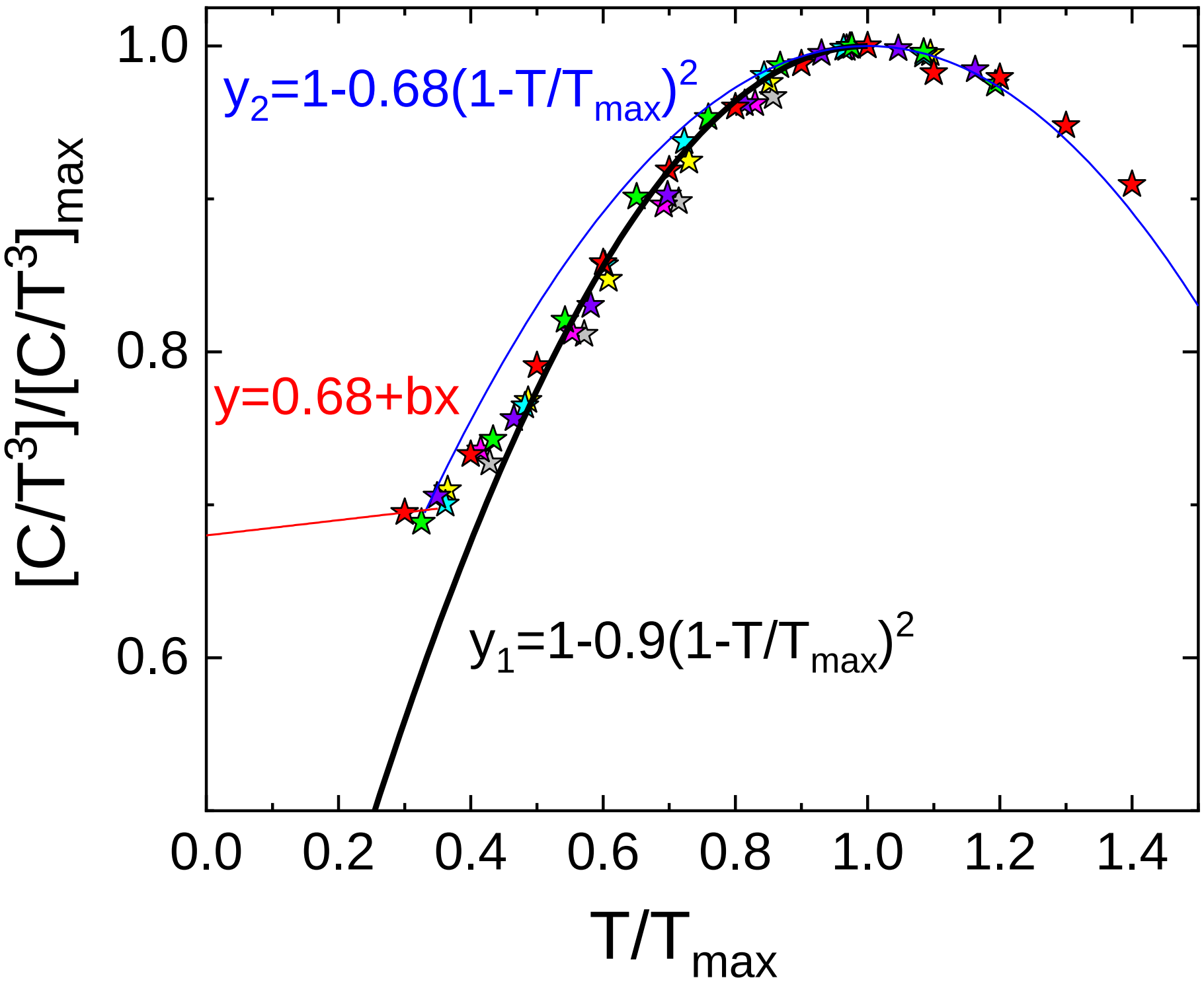

Fig. 11. Ratio $[C/T^3]/[C/T^3]_{\max}$ vs. $T/T_{\max}$ for $^3$He: ★ – $^3$He (V = 12.57 cm$^3$/mol) [34]; ★ – $^3$He $V$=13.33 cm$^3$/mol, [34]; ★ – $^3$He ($V$ = 13.56 cm$^3$/mol), [34]; ★ – $^3$He ($V$ = 14.11 cm$^3$/mol), [34]; ★ – $^3$He ($V$=14.16 cm$^3$/mol), [34]; ★ – $^3$He ($V$ = 14.98 cm$^3$/mol), [34]; ☆ – $^3$He ($V$ = 15.72 cm$^3$/mol), [34].

As in the case of other cryocrystals, for $^3$He at $T/T_{max}$<0.3 a linear dependence y=bx+$d$ is observed, where the intercept d agrees well with the experimental value $d_2=C_{Deb}/[C/T^3]_{max}\approx 0.68$. At higher reduced temperatures, $T/T_{max}>1$, the heat capacity for all studied $^3$He samples can be approximated by the quadratic function

$$y_2=1-d_2(1-T/T_{max})^2, \tag{11}$$

It is further seen that for $0.6<T/T_{max}$, the variation of molar volume does not affect the universal quadratic dependence near the hump, which can be expressed as

$$y_1=1-d_1(1-T/T_{max})^2, \tag{12}$$

where $d_1$=0.9 and the curve is noticeably asymmetric compared with $y_2$. The significant difference in the coefficients, $d_2\approx 0.68$, indicates that both quantum fluctuations and anharmonic effects play an essential role in determining the heat capacity of $^3$He.

The analysis of $^3$He demonstrates that the position and shape of the heat capacity hump are strongly influenced by molar volume and the interplay between van der Waals interactions and zero-point quantum fluctuations. The observed scaling relations and the distinct roles of the coefficients $d_1$ and $d_2$ confirm that quantum and anharmonic effects must be explicitly considered in light cryocrystals. These results provide a valuable benchmark for comparing with $^4$He and other rare-gas solids, thereby reinforcing the broader framework of universal thermodynamic behavior discussed in Section 3.

### 4.2. Desity effect on the hump in the heat capacity of the $^4$He

Figure 12 shows the temperature dependence of $C/T^3$ for a wide range of molar volumes of solid $^4$He: ■ – 11.77 cm³/mol, 12.22 cm³/mol, 14.55 cm³/mol, and 16.25 cm³/mol [34]; ● – 13.723 cm³/mol, 14.513 cm³/mol, 15.097 cm³/mol, 15.913 cm³/mol, 16.770 cm³/mol, 17.550 cm³/mol, 18.270 cm³/mol, and 19.135 cm³/mol [36]; ◆ – 16.90 cm³/mol, 17.87 cm³/mol, 18.22 cm³/mol, 19.18 cm³/mol, 19.68 cm³/mol, and 20.93 cm³/mol [35]; ▲ – 20.96 cm³/mol, 20.742 cm³/mol, 20.521 cm³/mol, and 20.456 cm³/mol [62]. An increase in molar volume $V$ leads to an increase in the absolute values of the heat capacity, reflecting changes in the phonon density of states associated with the lattice packing. A decrease of density (larger $V$) corresponds to a smaller effective (pseudo-) Brillouin zone [68 – 71], which facilitates the population of lower-energy vibrational states and enhances the specific heat at low temperatures. Furthermore, the maximum temperature, $T_{max}$ of the hump, systematically shifts to lower values with increasing $V$ (Fig. 12, Table 3).

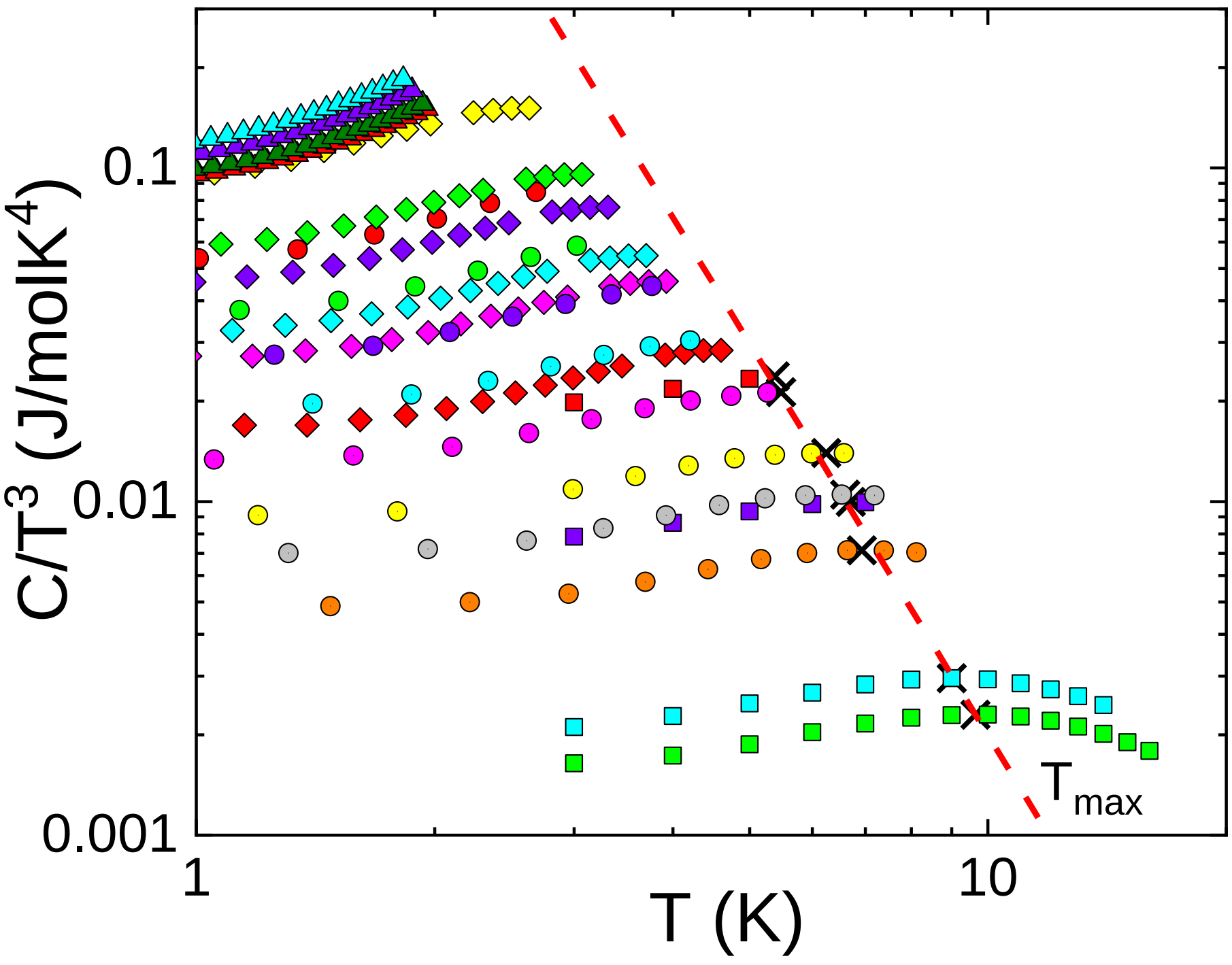

Fig. 12. Heat capacity of $C/T^3$ vs. $T$ for: ■ – $^4$He (V = 11.77 cm$^3$/mol) [34]; ■ – $^4$He (V = 12.22 cm$^3$/mol) [34]; ■ – $^4$He (V = 14.55 cm$^3$/mol) [34]; ■ – $^4$He (V = 16.25 cm$^3$/mol) [34]; ● – $^4$He (V = 13.723 cm$^3$/mol) [36]; ●- $^4$He (V = 14.513 cm$^3$/mol) [36]; ● – $^4$He (V = 15.097 cm$^3$/mol) [36]; ● – $^4$He (V = 15.913 cm$^3$/mol) [36]; ● – $^4$He (V = 16.770 cm$^3$/mol) [36]; ● – $^4$He (V = 17.550 cm$^3$/mol) [36]; ● – $^4$He (V = 18.270 cm$^3$/mol) [36]; ● – $^4$He (V = 19.135 cm$^3$/mol) [36]; ◆ – $^4$He (V = 20.93 cm$^3$/mol) [35]; ◆ – $^4$He (V = 19.68 cm$^3$/mol) [35]; ◆ – $^4$He (V = 19.18 cm$^3$/mol) [35]; ◆ – $^4$He (V = 18.22 cm$^3$/mol) [35]; ◆ – $^4$He (V = 17.87 cm$^3$/mol) [35]; ◆ – $^4$He (V = 16.90 cm$^3$/mol) [35] ▲ – $^4$He (V = 20.96 cm$^3$/mol) [62]; ▲ – $^4$He (V = 20.742 cm$^3$/mol) [62]; ▲ – $^4$He (V = 20.521 cm$^3$/mol) [62]; ▲ – $^4$He (V = 20.456 cm$^3$/mol) [62]; x – Estimated values for $T_{max}$ and line for the eye demonstrates the shift of the $T_{max}$.

Table 3. $T_{max}$, $C_{Deb}$ and $[C/T^3]_{max}$, $d=C_{Deb}/[C/T^3]_{max}$, for $^4$He. * – Estimated values for $T_{max}$ , for for $^4$He at $V > 15.913$ cm$^3$/mol. $[C/T^3]_{max}$ from the linear dependence of the $T_{max}$

| Material | $T_{max}$ | $C_{Deb,}$ mJ/molK4 | $[C/T^3]_{max,}$ mJ/molK4 | $C_{Deb}/[C/T^3]_{max}$ | Ref. |
|---|---|---|---|---|---|
| ▲ $^4$He (V = 20.96 cm$^3$/mol) | 2.87* | 109.58 | - | - | [62] |
| ▲ $^4$He (V = 20.742 cm$^3$/mol) | 2.93* | 101.48 | - | - | [62] |
| ▲ $^4$He (V = 20.521 cm$^3$/mol) | 3$^*$ | 92.53 | - | - | [62] |
| ▲ $^4$He (V = 20.456 cm$^3$/mol) | 3.02$^*$ | 89 | - | - | [62] |
| ◆ $^4$He (V = 20.93 cm$^3$/mol) Hexagonal Close-Packed | 2.88$^*$ | 89 | - | - | [35] |
| ◆ hcp $^4$He (V = 19.68 cm$^3$/mol) Hexagonal Close-Packed | 3.28$^*$ | 56.4 | - | - | [35] |
| ● $^4$He (V = 19.135 cm$^3$/mol) | 3.47$^*$ | 51.33 | - | - | [36] |
| ◆ hcp $^4$He (V = 19.18 cm$^3$/mol) Hexagonal Close-Packed | 3.46$^*$ | 44 | - | - | [36] |
| ● $^4$He (V = 18.270 cm$^3$/mol) [33_Ahlers_1970]; | 3.83$^*$ | 35,91 | - | - | [36] |
| ◆ hcp $^4$He (V = 18.22 cm$^3$/mol) | 3.85$^*$ | 32 | - | - | [35] |
| ◆ hcp $^4$He (V = 17.87 cm$^3$/mol) Hexagonal Close-Packed | 4.01$^*$ | 26.97 | - | - | [35] |
| ● $^4$He (V = 17.550 cm$^3$/mol) | 4.17$^*$ | 26.54 | - | - | [36] |
| ● $^4$He (V = 16.770 cm$^3$/mol) | 4.58$^*$ | 18.98 | - | - | [36] |
| ◆ hcp $^4$He (V = 16.90 cm$^3$/mol) Hexagonal Close-Packed | 4.51$^*$ | 16.5 | - | - | [35] |
| ■ 4He (V = 16.25 cm$^3$/mol) | 5.38 | 16 | 23.75 | 0.67 | [34] |
| ● $^4$He (V = 15.913 cm$^3$/mol) | 5.48 | 13 | 21.28 | 0.61 | [36] |
| ● $^4$He (V = 15.097 cm$^3$/mol) | 6.25 | 9 | 13.98 | 0.64 | [36] |
| ● $^4$He (V = 14.513 cm$^3$/mol) | 6.6 | 6.85 | 10.50 | 0.65 | [36] |
| ■ 4He (V = 14.55 cm$^3$/mol) | 6.71 | 6.8 | 9.98 | 0.68 | [34] |
| ● $^4He$ (V = 13.727 $cm^3$/mol) | 6.93 | 4.91 | 7.15 | 0.69 | [36] |
| ■ 4He (V = 12.22 $cm^3$/mol) | 9 | 2 | 2.92 | 0.69 | [34] |
| ■ $^4He$ (V = 11.77 $cm^3$/mol) | 9.64 | 1.56 | 2.3 | 0.68 | [34] |

The normalized ratio $[C/T^3]/[C/T^3]_{max}$ versus $T/T_{max}$ is presented in Figure 13. For $0.8 < T/T_{max}$, all studied $^4He$ samples exhibit a universal quadratic dependence $y_1 = 1 - d_1(1 - T/T_{max})^2$ near the hump. Deviations from this scaling appear for $T/T_{max} < 0.6$, while at $T/T_{max} < 0.2$, all cryocrystals show a linear dependence $y = ax + d$ with an intercept consistent with the Debye-based ratio $C_{Deb}/[C/T^3]_{max}$ (Table 3). At $T/T_{max} > 1$, the heat capacity is well approximated by $y_2 = 1 - d_2(1 - T/T_{max})^2$. The significant difference between the coefficients $d_1 = 0.85$ and $d_2 = 0.67$ indicates strong contributions from quantum and anharmonic effects in $^4He$. Importantly, increasing the molar volume results in a systematic decrease of the ratio $d = C_{Deb}/[C/T^3]_{max}$, from $d_2 = 0.67$ at V = 13.727 $cm^3$/mol [36] to $d_2 = 0.62$ at $V = 15.913$ $cm^3$/mol [36]. V.N. Grigor'ev and Ye.O. Vekhov [37] estimated that the relative vacancy contribution to the specific heat increases from $C_{vac}/C_{exp} \approx 0.05$ for hcp $^4He$ at $V = 13.727$ $cm^3$/mol to near 0.1 at V = 15.913 $cm^3$/mol. Thus, the measured $C_{exp}$ represents the sum of phonon and vacancy contributions, with the vacancy term becoming increasingly relevant at higher temperatures near $T_{max}$, while negligible in the $T/T_{max} \to 0$ limit. As a result, larger experimental values of $[C/T^3]_{max}$ compared to the phonon term $[C_{ph}/T^3]_{max}$ yield lower experimental estimates of $d = C_{Deb}/[C/T^3]_{max}$ than the purely phononic value $C_{Deb}/[C_{ph}/T^3]_{max}$.

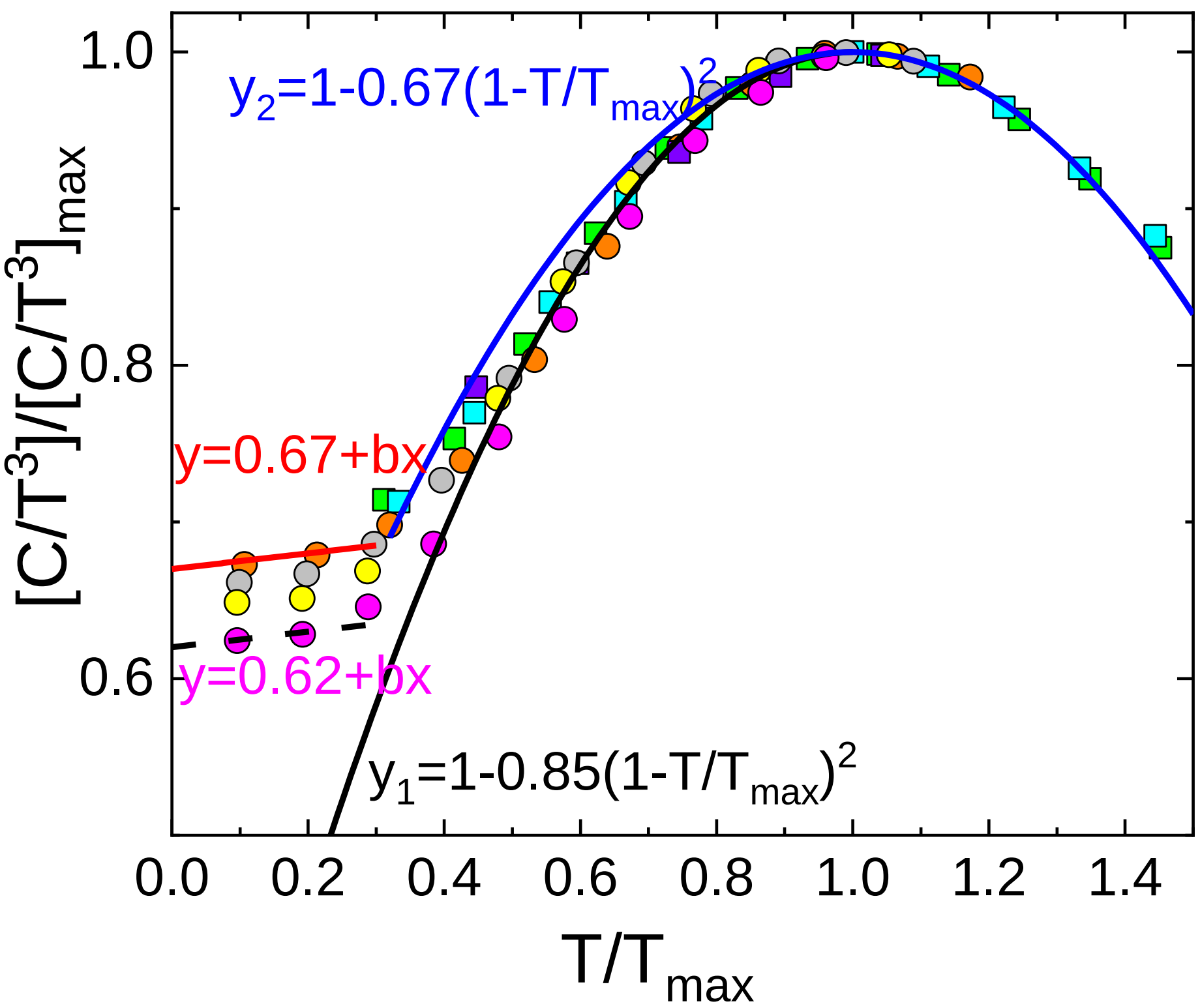

Fig. 13. Ratio $[C/T^3]/[C/T^3]_{max}$ vs. $T/T_{max}$ for: ■ – $^4$He (V = 11.77 cm$^3$/mol) [34]; ■ – $^4$He (V = 12.22 cm$^3$/mol) [34]; ■- $^4$He (V = 14.55 cm$^3$/mol) [34]; ■ – $^4$He (V = 16.25 cm$^3$/mol) [24]; ● – $^4$He (V = 13.723 cm$^3$/mol) [36]; ● – $^4$He (V = 14.513 cm$^3$/mol) [36]; ● – $^4$He (V = 15.097 cm$^3$/mol) [36]; ● – $^4$He (V = 15.913 cm$^3$/mol) [36];

The character of the $\Delta^*$ dependence is further analyzed in Figure 14. For $T < T_{max}$, the data for Ne, $^3$He, and $^4$He follow a nearly universal relation $\Delta^* = 1 - 3(1 - T/T_{max})^2$. Increasing the molar volume of $^4$He from 12.22 cm³/mol [34] to 15.913 cm³/mol [36] does not affect this branch. For Ne, symmetrical dependencies are observed both below and above $T_{max}$, whereas for quantum solids $^3$He and $^4$He, $\Delta^*$ is asymmetrical. For example, ■ – $^4$He (V = 12.22 cm³/mol) [34] and – $^3$He (V = 12.57 cm³/mol) [34] exhibit asymmetrical branches characterized by $\Delta^* = 1 - 1.7(1 - T/T_{max})^2$ at $T > T_{max}$. With increasing molar volume, the melting temperature of $^4$He decreases, preventing observation of the second asymmetrical branch in high-V samples such as ● – $^4$He (V = 15.913 cm³/mol) [36]. These asymmetrical forms of $\Delta^*$ above and below $T_{max}$ highlight the strong anharmonicity of helium crystals and its impact on phonon vibrations.

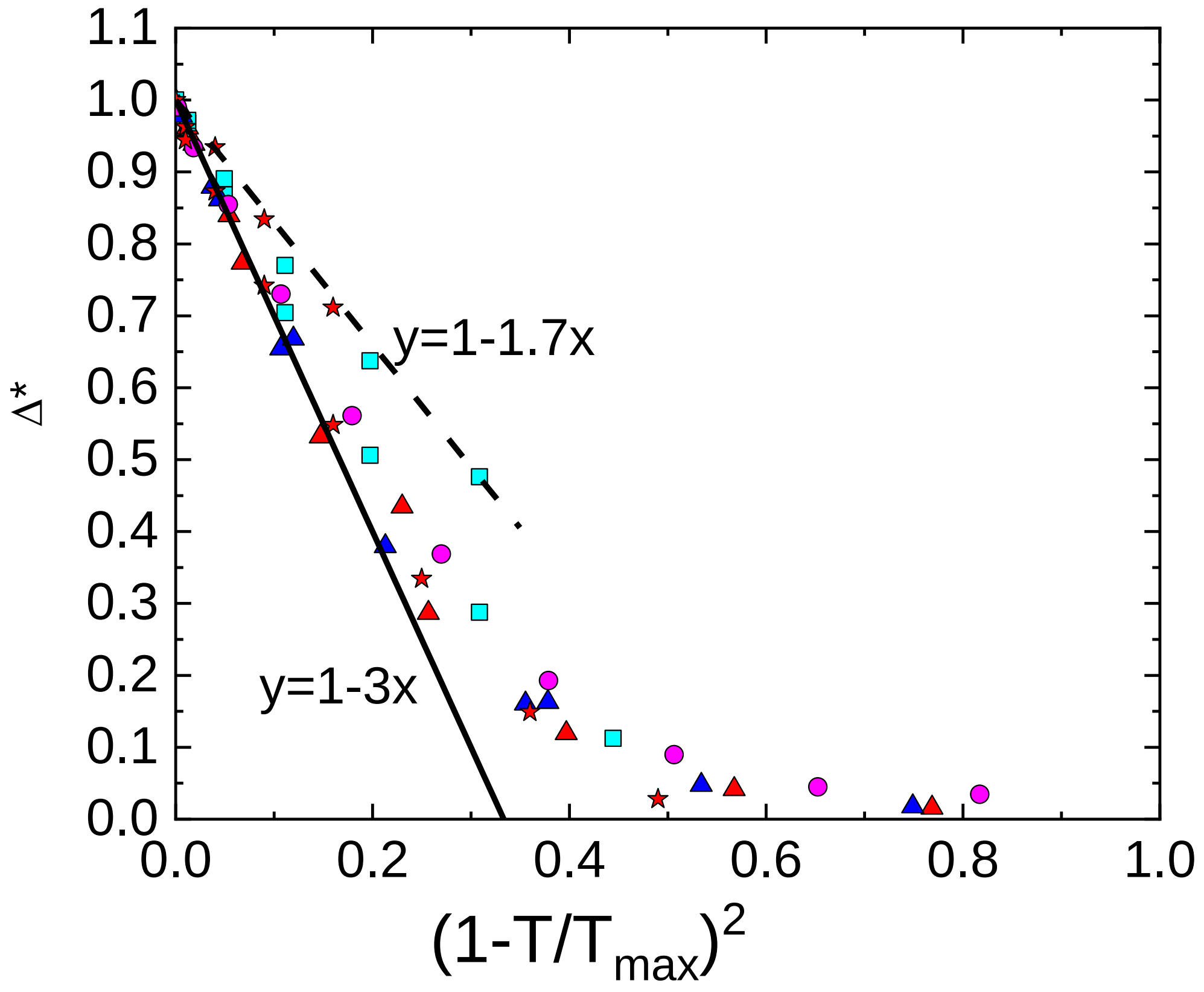

Fig. 14. $\Delta^*$ vs. $(1-T/T_{max})^2$ for: ■ – $^4$He ($V = 12.22$ cm$^3$/mol) [34]; ★ – $^3$He ($V = 12.57$ cm$^3$/mol) [34]; ● – $^4$He ($V = 15.913$ cm$^3$/mol) [36]; ▲ – Ne ($V = 12.39$ cm$^3$/mol) [39]; ▲ – Ne ($V = 12.87$ cm$^3$/mol) [39];

Overall, the ratio $[C/T^3]/[C/T^3]_{max}$ vs. $T/T_{max}$ and the $\Delta^*$ vs. $(1 - T/T_{max})^2$ scaling relations serve as sensitive probes of anharmonicity, quantum effects, vacancy contributions, and lattice disorder in helium solids. This behavior of $^4$He, while sharing the universal features observed in rare-gas cryocrystals, also parallels the asymmetry and quantum-driven deviations identified in $^3$He, underlining the unique interplay of density, anharmonicity, and quantum effects in helium solids.

The density dependence of the heat capacity hump in $^4$He reveals both universal scaling features and clear deviations driven by quantum and anharmonic effects. The systematic reduction of the parameter $d$ with increasing molar volume, together with the growing vacancy contribution, demonstrates that phonon-vacancy interactions play a crucial role near $T_{max}$. Compared with Ne and Ar, the asymmetrical $\Delta^*$ branches of $^4$He highlight its stronger anharmonicity and its closer resemblance to the quantum behavior of $^3$He.

### 4.3. Desity effect on the hump in the heat capacity of the $H_2$

The constant-volume heat capacity of solid parahydrogen (p-$H_2$) has been measured from 4 K to the melting line for six molar volumes (22.787 to 16.19 cm$^3$/mole) by Krause [28]. As well as for helium, for the light elements $H_2$, $D_2$ the constant volume conditions are important at low temperatures due to the high difference $C_p$-$C_V$ between values of heat capacity at constant pressure $C_p$ and volume $C_V$. In addition, the zero point energy amounts are approximately 47% of the cohesion energy at 0°K for hydrogen [72], which is much higher than in the case of rare gas solids of argon with 8% of this figure [73].

Let's analyze the influence of the change of molar volume on the character $\Delta^*$ vs. $T/T_{max}$ for hydrogen $H_2$ [28]. It can be seen in Fig. 15, that the change in the molar volume for hydrogen $H_2$ does not affect the character of $\Delta^*$ vs. $T/T_{max}$ in the range $0.6 < T/T_{max} < 1.2$ as well as in the case of classical cryocrystals Ne. It can be well described by a quadratic dependence (eq.11) with D=3:

$$\Delta^* = 1 - D_{left}(1 - T/T_{max})^2, \tag{13}$$

Isoterms $T/T_{max}$= 1.4 and $T/T_{max}$= 2 are shown by dot lines for the eye on the figures. It is seen, that at $T > T_{max}$ the value $\Delta^*$ at the same temperature is higher for the cryocrystal of $H_2$ with higher molar volume that agrees with the decrease of the ratio $d = C_{Deb}/[C/T^3]_{max}$ (Table 4). As in the case of the $^4$He, it indicates the increase of influence anharmonicity or vacancy due to the decrease of the melting (triple point) temperature with the increase of the molar volume. For ▲ – $H_2$ (V = 16.193 cm$^3$/mol) [28] the experimental data for $\Delta^*$ are available at the interval of $T/T_{max}$ from 0 to 2.4. For ■ – $H_2$ (V = 22.221 cm$^3$/mol) [28] the experimental data for $\Delta^*$ are available at the interval of $T/T_{max}$ from 0 to 1.4.

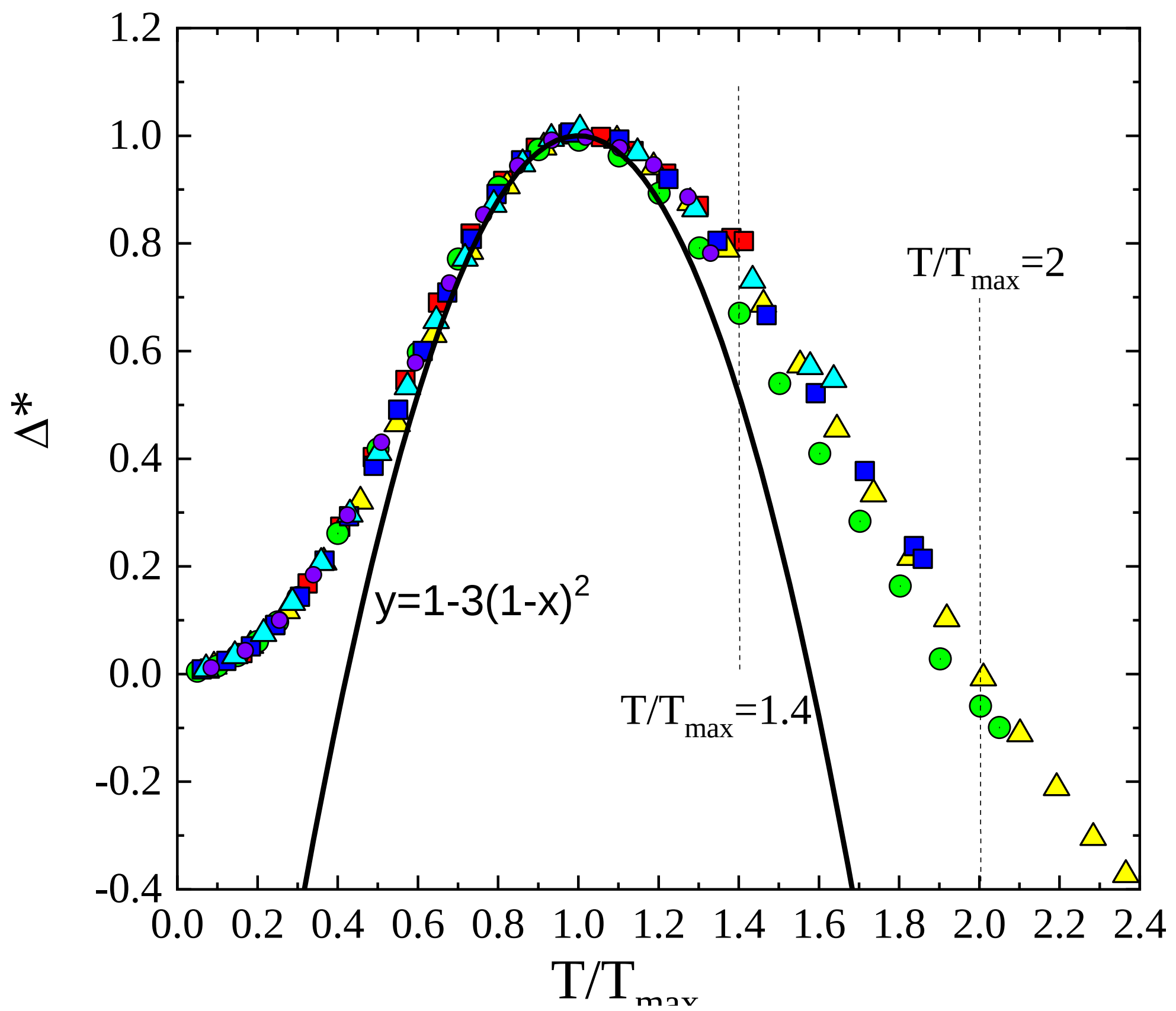

Fig. 15 $\Delta^*$ vs. $T/T_{max}$: ▲ – $H_2$ (V = 16.193 cm$^3$/mol) [28]; ● – $H_2$ (V = 17.458 cm$^3$/mol) [28]; ■ – $H_2$ (V = 19.12 cm$^3$/mol) [28]; ▲ – $H_2$ (V = 20.685 cm$^3$/mol) [28]; ■ – $H_2$ (V = 22.221 cm$^3$/mol) [28] ● – $H_2$ (V = 22.787 cm$^3$/mol) [28].

Table 4. $T_{max}$, $C_{Deb}$ and $[C/T^3]_{max,}$ d=$C_{Deb}$ /$[C/T^3]_{max,}$ $D_{left}$ for $H_2$.

| Material | $T_{max}$ | $C_{Deb,exp}$ mJ/molK4 | $[C/T^3]_{max,}$ mJ/molK4 | *d* | *D* | Ref. |
|---|---|---|---|---|---|---|
| S6, $H_2$, V=16.193 $cm^3$/mol | 21.89 | 0.12 | 0.18 | 0.64 | 3 | [28]. |
| S5, $H_2$, V=17.458 $cm^3$/mol | 19.97 | 0.18 | 0.28 | 0.64 | 3 | [28]. |
| S3, $H_2$, V=19.12 $cm^3$/mol | 16.33 | 0.31 | 0.48 | 0.64 | 3 | [28]. |
| s4, $H_2$, V=20.685 $cm^3$/mol | 13.94 | 0.51 | 0.81 | 0.63 | 3 | [28]. |
| S1, $H_2$, V=22.221 $cm^3$/mol | 12.3 | 0.8 | 1.31 | 0.62 | 3 | [28]. |
| $H_2$, (V = 22.787 $cm^3$/mol) | 11.66 | 0.97 | 1.56 | 0.62 | 3 | [28]. |

Normalized dimensionless function $\Delta^*(T)$ for specific heats of two cryocrystals with different molar volumes are the same at the same values of $T/T_{max}$ if their respective dimensionless phonon frequency distribution functions $g(\omega)\omega^2=G(\omega/\omega_D)$ are the same function of $\omega/\omega_D$ [15]. From the obtained universally dependence $\Delta^*$vs. $T/T_{max}$ for all cryocrystals in the range from $0<T/T_{max}<1.2$ and symmetrical character of this dependence for clasical cryocrystals (Ar, Kr, Ne, Xe) and asymmetrical for the quantum cryocrystals ($^3$He, $^4$He, $H_2$), it is possible to assume that in the temperature range where anharmonicity to the heat capacity can be neglected, the normalized phonon spectrum of crystals can be written as a function of the frequency $\omega_{vH}$ of the first van Hove singularity:

$$g(\omega)/\omega^2 = F(\omega/\omega_{vH}) \quad , \qquad (15)$$

This leads to the universality of the behavior of the heat capacity $\Delta^*$ and the appearance of peculiarities at the temperature $T_{max}$:

$$\frac{C(T)}{T^3} = G\left(\frac{T}{\Theta_{vH}}\right) \ , \tag{16}$$

, where $\Theta_{vH} \simeq 5T_{max}$ [16].

## 5. Conclusions

In this work, we have systematically analyzed the heat capacity behavior of classical rare-gas cryocrystals (Xe, Kr, Ar, Ne) and quantum cryocrystals ($^3$He, $^4$He, $H_2$) under variations of molar volume and external pressure. A universal quadratic scaling of the reduced heat capacity near the hump temperature $T_{max}$ has been established for all systems, with the coefficient D=3 capturing the effect of the van Hove singularity shift under lattice compression or expansion. For classical cryocrystals, this scaling is symmetric with respect to $T_{max}$, reflecting the dominance of phonon density-of-states effects and a nearly harmonic lattice response.

For quantum crystals ($^4$He, $^3$He, $H_2$), as in the case of classical cryocrystals (Ar, Kr, Xe), an anomaly in the heat capacity $C/T^3$ vs. $T$ with a maximum at $T_{max}$ is observed. The dependence of the anomaly parameters $T_{max}$, $[C/T^3]_{max}$, and $C_{Deb}$ on molar volume reveals a proportional relationship between these quantities and volume. This proportionality indicates that changes in the first Brillouin zone with molar volume affect both the shift of the first van Hove frequency $\omega_{vH}$ and the Debye cutoff frequency $\omega_D$. Thus, the observed scaling confirms that lattice compression and expansion govern phonon spectra in a unified way across both classical and quantum cryocrystals.

In contrast, the quantum cryocrystals also exhibit pronounced deviations from symmetry due to strong zero-point fluctuations, anharmonic effects, and thermal vacancies, which significantly modify the shape of the heat capacity hump and reduce the ratio $d=C_{Deb}/[C/T^3]_{max}$. These effects become more pronounced at larger molar volumes, where lower melting temperatures enhance vacancy contributions. Similar asymmetries are observed in solid parahydrogen, where the exceptionally high zero-point energy ($\approx$47% of the cohesion energy) amplifies the role of anharmonicity compared with the classical rare gases.

Taken together, these results provide a coherent comparative framework for interpreting heat capacity anomalies in cryocrystals. The scaling representations $[C/T^3]/[C/T^3]_{max}$ vs. $T/T_{max}$ and $\Delta^*$ vs. $(1-T/T_{max})^2$ emerge as powerful tools for disentangling harmonic phonon contributions from quantum and defect-induced effects. The demonstrated universality of $\Delta^*$ across both quantum ($^4$He, $^3$He, $H_2$) and classical (Ar, Kr, Xe) cryocrystals emphasizes the fundamental role of lattice dynamics in governing low-temperature thermodynamic behavior.

**Funding:**

This work was partly supported by the National Research Foundation of Ukraine (Grant 2023.03/0012) and National Science Centre Poland (Grant 2022/45/B/ST3/02326).

## Ref.